\documentclass[preprint,12pt,review]{elsarticle}
\usepackage[pass]{geometry} 

\usepackage{kotex}
\usepackage{amsmath}
\usepackage{amssymb}
\usepackage{lineno,hyperref}
\usepackage{subcaption}
\usepackage{xcolor}
\usepackage{graphicx}
\journal{Aerospace Science and Technology}
\biboptions{sort&compress}
\usepackage{afterpage}
\usepackage{gensymb}
\usepackage{pdflscape}  
\usepackage{adjustbox}
\usepackage{afterpage}
\usepackage{multirow}
\usepackage{array}

\modulolinenumbers[5]
\begin{document}

\begin{frontmatter}
    

\title{Effects of Surface Corrugation on Gas-Surface Scattering and Macroscopic Flows}
\author[inst1]{Woonghwi Park}
\author[inst1]{Eunji Jun\corref{cor}}
\affiliation[inst1]{organization={Korea Advanced Institute of Science and Technology},
            postcode ={34141}, 
            state={Daejeon},
            country={Republic of Korea}}

\ead{eunji.jun@kaist.ac.kr}
\cortext[cor]{Corresponding author}

\date{\today}


\begin{abstract}
Gas–surface scattering exhibits a transition from thermal to structure scattering regime as incident energy increases, characterized by changes in angular and energy distributions. Capturing scattering behavior across different regimes is essential for modeling momentum and energy exchange at the gas–surface interface. This study presents a corrugated Cercignani–Lampis–Lord (CLL) model to account for surface corrugation in scattering. It extends the washboard–CLL hybrid approach by incorporating tangential momentum accommodation in local collisions. The model is validated against molecular beam scattering experiments, focusing on its ability to reproduce the variation in scattering behavior with increasing incident energy. Compared to the conventional CLL model, the proposed model qualitatively improves the representation of key features across the scattering regimes. In particular, it captures broader angular distributions and increasing reflected energy with reflected angle at higher incident energy. It is further applied to Direct Simulation Monte-Carlo (DSMC) analysis of rarefied flows over a cylinder and within an intake to examine the influence of surface corrugation on macroscopic flow behavior. The results show that surface corrugation has limited impact on total drag, while decreasing pressure drag and increasing friction drag. In the intake, enhanced tangential accommodation reduces capture efficiency and increases compression ratio. These findings highlight the importance of incorporating surface corrugation in rarefied flow simulations under VLEO conditions.

\end{abstract}

\end{frontmatter}

\section{Introduction} \label{sec:1}

Gas-surface scattering models describe collisions between gas particles and solid surfaces, characterizing momentum and energy exchange at the interface. Molecular beam scattering experiments study the trajectories of gas particles upon collision with surfaces to improve understanding of gas-surface scattering~\cite{bernasek1975molecular,oman1968numerical,smith1973molecular,rettner1991angular,livadiotti2020review}. These studies reveal that as incident energy increases, a transition occurs from the thermal scattering regime to the structure scattering regime. In the thermal scattering regime, low incident energies prevent gas particles from penetrating the surface potential, resulting in energy exchange dominated by surface thermal motion~\cite{weinberg1975molecular,somorjai1973interaction}. In the structure scattering regime, higher incident energies allow gas particles to penetrate the surface potential, shifting the dominant scattering mechanism from surface thermal motion to surface corrugation~\cite{amirav1987atom,smith1973molecular,deng2022modified,ruiling2024gas}. A scattering model that incorporates both surface thermal motion and surface corrugation is required to capture thermal and structure scattering.

The Direct Simulation Monte Carlo (DSMC) method, a particle-based kinetic approach, is widely used to analyze rarefied flows~\cite{bird1994molecular,bird1998recent}. In this method, statistical kernels such as the Maxwell and Cercignani–Lampis–Lord (CLL) models are used to describe gas–surface scattering based on full or partial accommodation to surface thermal energy~\cite{maxwell1890scientific,cercignani1971kinetic}. By accounting for surface temperature, the Maxwell and CLL models capture thermal scattering. These models also satisfy reciprocity, ensuring microscopic reversibility and thermodynamic consistency~\cite{wenaas1971equilibrium,cercignani1972scattering}. However, these models do not account for surface corrugation. Although previous studies have fitted the CLL model to reproduce structure scattering, such approaches do not reflect the underlying physics of structure scattering~\cite{liu2021dsmc}.

The washboard model introduces surface corrugation into the modeling of gas–surface scattering~\cite{tully1990washboard,yan2004washboard,mateljevic2009accommodation}. It extends the hard-cube (HC) model by incorporating surface corrugation through a sinusoidal profile~\cite{goodman1965theory,logan1966simple,logan1968classical}. While the HC model assumes impulsive collisions with surface atoms vibrating normal to a flat surface, the washboard model replaces this direction with the local normal defined by the sinusoidal profile. Other assumptions of the HC model, including momentum exchange limited to the normal direction and conservation of tangential momentum, are retained. The washboard model successfully reproduces the rainbow scattering, the bimodal angular distribution observed in the structure scattering regime~\cite{tully1990washboard}. However, its failure to satisfy reciprocity limits its direct applicability in DSMC~\cite{liang2018physical}.

A hybrid approach combining the washboard and CLL models has been developed to incorporate surface corrugation while preserving reciprocity~\cite{liang2018physical,liang2021parameter}. In this approach, the HC model in the original washboard model is replaced by the CLL model. The assumption of tangential momentum conservation during collisions is retained. Collisions in shadowed regions and multiple interactions arising from surface corrugation are treated using a re-collision probability, which ensures that the model satisfies reciprocity. This hybrid approach captures key features of microscopic scattering, including the dependence of momentum and energy accommodation on incident energy. However, the application of this hybrid approach in DSMC for studying surface corrugation effects on macroscopic flows has not been adequately explored.

This study investigates the effects of surface corrugation on gas-surface scattering and its influence on macroscopic flows. A modified hybrid model, referred to as the corrugated CLL model, is proposed in this study. It extends the washboard–CLL hybrid approach by incorporating tangential momentum accommodation that may arise during local gas-surface collisions. The proposed model is validated by reproducing angular and energy distributions from molecular beam scattering experiments, focusing on the model's ability to capture the transition from thermal to structure scattering regimes~\cite{rettner1991angular}. The corrugated CLL model is then applied to analyze the impact of surface corrugation on macroscopic flows around a cylinder and within an intake. The remainder of this paper is structured as follows. Section~\ref{sec:2} provides an overview of scattering models. The formulation of the corrugated CLL model and its integration into the DSMC framework are described in Section~\ref{sec:3}. Numerical results from molecular beam scattering simulations are reported in Section~\ref{sec:4}, followed by an analysis of macroscopic flow behavior under surface corrugation effects in Section~\ref{sec:5}. Finally, Section~\ref{sec:6} summarizes the findings and outlines directions for future work.

\clearpage

\afterpage{%
\footnotesize
\newgeometry{top=2.5cm, bottom=2.5cm, left=2.5cm, right=2.5cm}

    \clearpage
    \thispagestyle{empty}
    \begin{landscape}
        \centering 
        \captionof{table}{Scattering Models}
            \label{tab:scattering_model}

\begin{tabular}{
>{\raggedright\arraybackslash}p{2.5cm} 
>{\raggedright\arraybackslash}p{2.5cm} 
>{\raggedright\arraybackslash}p{6cm} 
>{\raggedright\arraybackslash}p{5cm} 
>{\raggedright\arraybackslash}p{5cm}}            \hline
\textbf{Category} & \textbf{Model} & \textbf{Reflection Mechanism} & \textbf{Advantages} & \textbf{Limitations} \\ \hline

\multirow{2}{*}{\parbox[c]{2cm}{\textbf{Scattering Kernels}}}  & \textbf{Maxwell~\cite{maxwell1890scientific}} & 
Linear combination of diffuse and specular reflections using $\sigma$
& 
Simple formulation; reciprocity satisfied
&
Inability to capture lobular angular distributions \\ \cline{2-5}
&\textbf{CLL~\cite{cercignani1971kinetic}} &
Independent accommodation in normal and tangential directions using $\sigma_n$, $\sigma_t$
&
Reproduction of lobular angular distribution; reciprocity satisfied
&
Complex and nontrivial parameter selection
\\ \hline

\multirow{2}{*}{\parbox[c]{2cm}{ \textbf{Physical Models}}} 
            & \textbf{Hard~Cube (HC)~\cite{goodman1965theory}} & 
Impulsive collision between rigid and elastic bodies with no tangential friction
& 
Representation of thermal scattering on smooth surfaces
&
Neglect of attractive forces; violation of reciprocity \\ \cline{2-5}

& \textbf{Washboard~\cite{tully1990washboard}} &
Sinusoidal surface profile with local collisions based on HC model along tilted surface normals
&
Inclusion of surface corrugation; rainbow scattering prediction
&
Violation of reciprocity; incompatibility with DSMC
\\ \hline

\textbf{Hybrid Approach} 
 & \textbf{Washboard-CLL Hybrid~\cite{liang2018physical}} & CLL-based local scattering applied on a 3D random Gaussian hills and valleys, with probabilistic re-collision&Inclusion of surface corrugation; reciprocity satisfied through re-collision mechanism & Neglect of tangential force at the local collision surface
\\ \hline
        \end{tabular}

    \end{landscape}
    \clearpage
}
\restoregeometry

\clearpage

\section{Review of Scattering Models}\label{sec:2}
\subsection{Scattering Kernels}

The scattering kernel $K(\mathbf{v}_r|\mathbf{v}_i)$ represents the probability density function for obtaining a reflected velocity $\mathbf{v}_r$ given an incident velocity $\mathbf{v}_i$. The Maxwell and CLL models are commonly used to define $K(\mathbf{v}_r|\mathbf{v}_i)$, each offering a different approach to modeling energy and momentum exchange at the interface.

\subsubsection{Maxwell Model}
The Maxwell model assumes a fraction $\sigma$ of incident particles undergo diffuse reflection, while the remaining $1-\sigma$ reflect specularly~\cite{maxwell1890scientific}. Diffuse reflection involves complete thermal accommodation to the surface. Specular reflection is an elastic reflection with no energy exchange, symmetric about the surface normal. The energy accommodation coefficient $\sigma$ quantifies the degree of thermal equilibrium with the surface and is defined as: 
\begin{equation}
    \sigma = \frac{\langle \frac{1}{2} m_g v_{i}^2 \rangle- \langle \frac{1}{2} m_g v_{r}^2 \rangle} {\langle \frac{1}{2} m_g v_{i}^2 \rangle - E_{w}}.
    \label{eq:eac}
\end{equation}
Here, $m_g$ is the mass of gas particle, $v_{i}$ and $v_{r}$ denote magnitudes of the incident and reflected velocities, $E_{w}=\frac{3}{2}k_B T_w$ represents the energy of a gas particle equilibrated at the wall temperature $T_w$, and $k_B$ is the Boltzmann constant $1.38\times 10^{-23}$~J/K. The Maxwell model's scattering kernel is formulated as:
\begin{equation}
    K_{\text{Maxwell}}(\mathbf{v}_r|\mathbf{v}_i) = \sigma K_{\text{diffuse}}(\mathbf{v}_r) + (1-\sigma) K_{\text{specular}} (\mathbf{v}_r|\mathbf{v}_i).
\end{equation}
In this expression, the diffuse component $K_{\text{diffuse}}$ follows a Maxwell–Boltzmann distribution at $T_w$ and the specular component $K_{\text{specular}}$ follows a Dirac delta function representing elastic reflection with respect to the surface normal. Two limiting cases are defined by: $\sigma=1$ indicates complete thermalization of gas particles with the surface, while $\sigma=0$ corresponds to purely specular reflection. Due to its simple formulation, the Maxwell model is widely used as a boundary condition in many DSMC studies~\cite{moon2023performance,pfeiffer2018particle}. However, it fails to capture realistic lobular angular distributions observed in molecular beam scattering experiments~\cite{murray2017gas}.

\subsubsection{Cercignani–Lampis–Lord Model}
The CLL model introduces anisotropic accommodation in the normal and tangential directions~\cite{cercignani1971kinetic}. The normal and tangential energy accommodation coefficients, $\sigma_n$ and $\sigma_t$, are defined as:
\begin{align}
\sigma_n = \frac{\langle \frac{1}{2} m_g v_{i,n}^2 \rangle- \langle \frac{1}{2} m_g v_{r,n}^2 \rangle} {\langle \frac{1}{2} m_g v_{i,n}^2 \rangle - E_{w,n}},   \\
\sigma_t = \frac{\langle \frac{1}{2} m_g v_{i,t}^2 \rangle- \langle \frac{1}{2} m_g v_{r,t}^2 \rangle} {\langle \frac{1}{2} m_g v_{i,t}^2 \rangle - E_{w,t}},  
\label{eq:nteac}
\end{align}
where subscripts $n$ and $t$ correspond to the normal and tangential components, respectively.
The scattering kernel of the CLL model is given by:
\begin{align}
K_{\text{CLL}}&(\mathbf{v}_r|\mathbf{v}_i) = \sqrt{\frac{m_g^3}{2\pi k_B^3 T_w^3}} \frac{v_{r,n}}{\sigma_n \sqrt{\sigma_t}} \exp{\left(- \frac{m_g}{2k_B T_w} \frac{\left( v_{r,t} - \sqrt{1-\sigma_t}v_{i,t}\right)^2}{\sigma_t} \right)} \notag \\
& \times \exp {\left( - \frac{m_g}{2k_B T_w} \frac{v_{r,n}^2 + (1-\sigma_n) v_{i,n}^2}{\sigma_n}\right)} I_0 \left( \frac{m_g}{k_B T_w} \frac{ \sqrt{1-\sigma_n} v_{r,n}v_{i,n}}{\sigma_n} \right)
\label{eq:cll}
\end{align}
where $I_0$ is the modified Bessel function of the first kind~\cite{padilla2009assessment}. The CLL model captures the lobular angular distribution observed in molecular beam scattering experiments, which is not represented by the Maxwell model. However, parameter selection is nontrivial, as both $\sigma_n$, $\sigma_t$ must be simultaneously optimized to match experimental results~\cite{yousefi2019determination}.

\subsection{Physical Models}
Physical models provide a simplified classical-mechanical description of gas–surface scattering to elucidate its underlying dynamics. Two representative physical models are introduced: the HC model, which considers interactions with a single surface atom, and the washboard model, which incorporates surface corrugation.

\subsubsection{Hard Cube Model}
In the HC model, the surface is modeled as a lattice of atoms, each treated as a rigid cube~\cite{goodman1965theory}. Each atom vibrates perpendicular to the surface with a Maxwellian velocity distribution determined by the surface temperature. A gas particle undergoes an elastic and impulsive collision with a surface atom. During collisions, tangential friction is neglected, and the gas particle's tangential velocity is conserved. The normal velocity of the reflected gas particle is determined by the pre-collision velocities of the gas particle and the surface atom, as well as their mass ratio. The HC model qualitatively reproduces the thermal scattering on smooth surfaces~\cite{livadiotti2020review}. However, the assumptions of impulsive collisions and negligible attractive interactions restrict its applicability to adsorptive phenomena. Moreover, the HC model violates reciprocity because the reflected velocity distribution explicitly depends on the incident velocity direction, breaking symmetry under velocity reversal~\cite{goodman1971review}.

\subsubsection{Washboard Model and Its Extensions}

Originally proposed by J.C. Tully, the washboard model is formulated with two main components: surface geometry and local collision dynamics~\cite{tully1990washboard}. The surface geometry is defined by a two-dimensional sinusoidal profile, which introduces spatially varying surface normals that characterize corrugation. Collisions are resolved in the coordinate system defined by this local surface normal. For the collision dynamics, gas particles interact with thermally vibrating surface atoms along this local normal direction, following the HC model framework. An attractive potential can optionally be included to represent trapping-desorption processes.


Mateljevic et al. extended the surface representation in the original washboard model to three dimensions by defining the local surface normal using a polar angle $\alpha$ and an azimuthal angle $\beta$ relative to the surface normal~\cite{mateljevic2009accommodation}. Instead of a sinusoidal surface, this formulation employs a random geometry composed of Gaussian-shaped hills and valleys. The polar angle $\alpha$ follows a Rayleigh-like distribution,
\begin{equation}
f(\alpha,\beta) = \frac{1}{4A^2}\frac{\sin\alpha}{\cos^3\alpha}\exp\left(-\frac{\pi}{4A^2}(\sec^2\alpha-1)\right),
\label{eq:polarangle}
\end{equation}
for $0\le\alpha\le\pi/2$, while the azimuthal angle $\beta$ is uniformly distributed over $[0,2\pi]$. The corrugation factor $A$,
\begin{equation}
A = \langle \tan \alpha \rangle,
\end{equation}
represents the average slope of the surface. In addition, determining the probability distribution of the local surface normal at collision requires consideration of the particle's incident direction. For instance, surface elements whose local normals face toward the incident trajectory have higher collision probabilities than those oriented away from the trajectory. Given the particle's incident angle $\theta_i$, the conditional probability distribution for the local normal orientation becomes:
\begin{equation}
P_{\text{normal}}(\alpha,\beta;A,\theta_i) = f(\alpha,\beta)\max\left[\sin\theta_i\tan\alpha\cos\beta + \cos\theta_i,\,0\right].
\label{eq:tiltangle}
\end{equation}
This expression accounts for the projected area of the surface in the direction of incidence.

\subsection{Hybrid Approach}
\subsubsection{Washboard-CLL Hybrid Approach}
Liang et al. proposed a hybrid approach combining the washboard and CLL models, which incorporates surface corrugation while preserving reciprocity~\cite{liang2018physical,liang2021parameter}. The surface geometry follows the three-dimensional extension introduced by Mateljevic et al.~\cite{mateljevic2009accommodation}. In the local collision process, the CLL model replaces the HC model while maintaining conservation of tangential momentum. To ensure trajectory reversibility, reflected particles undergo probabilistic re-collision. The re-collision probability accounts for the possibility that a reflected particle may either be blocked by the geometry of the surface corrugation or undergoes a secondary collision with the surface. The re-collision probability $P_{\text{re-coll}}$ is defined as:
\begin{equation}
P_{\text{re-coll}}(\theta_r,\phi_r;\alpha,\beta) = \frac{f(\alpha,\beta)\,\max\left[-\left(1+\tan\theta_r\tan\alpha\cos(\phi_r-\beta)\right),0\right]}{\int f(\alpha,\beta)\,\max\left[1+\tan\theta_r\tan\alpha\cos(\phi_r-\beta),0\right]\,d\alpha\,d\beta},
\label{eq:reciprocity}
\end{equation}
where $\theta_r$ and $\phi_r$ denote the polar and azimuthal angles of the reflected velocity. By incorporating the re-collision probability, the model enforces reciprocity even in the presence of surface corrugation. An attractive potential is additionally incorporated to account for trapping–desorption. Particles reflected with insufficient velocity are drawn back toward the surface by this potential, resulting in repeated collisions. These repeated interactions enhance equilibration with the surface, thereby providing an indirect representation of the trapping–desorption process. The scattering models discussed above, including scattering kernels, physical models, and the hybrid approach, are summarized in Table~\ref{tab:scattering_model}.


\section{Corrugated CLL Model and Numerical Implementation}\label{sec:3}

\subsection{Model Description}
The corrugated CLL model builds on the washboard-CLL hybrid approach proposed by Liang et al. but incorporates tangential momentum accommodation in the local collision~\cite{mateljevic2009accommodation,liang2018physical,liang2021parameter}. Surface corrugation is represented by a three-dimensional random distribution of hills and valleys, as given in Eq.~\eqref{eq:polarangle}~\cite{mateljevic2009accommodation}. Local collisions are modeled using the CLL model given in Eq.~\eqref{eq:cll}, and tangential accommodation is included to account for imperfect conservation of tangential momentum. The re-collision probability given by Eq.~\eqref{eq:reciprocity} is applied to ensure reciprocity. The attractive potential used in the hybrid approach by Liang et al. is omitted in the corrugated CLL model to minimize trapping-desorption effects~\cite{liang2018physical}.

\subsection{Numerical Implementation}

\begin{figure}
    \centering
    \includegraphics[width=0.9\linewidth]{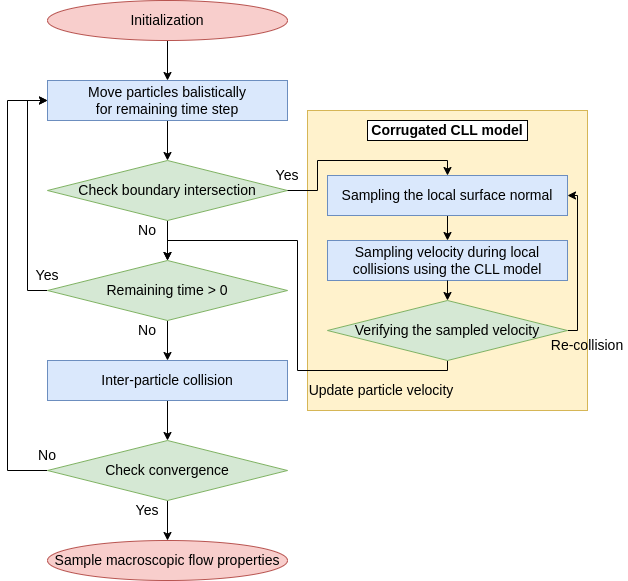}
    \caption{Flowchart of the corrugated CLL model integrated into DSMC}
    \label{fig:flowchart}
\end{figure}

In this study, the DSMC method is employed to analyze macroscopic flows. The DSMC method stochastically solves the Boltzmann equation by simulating the microscopic motion of gas particles~\cite{bird1994molecular,bird1998recent}. The open-source DSMC solver SPARTA is used~\cite{plimpton2019direct}. The overall algorithm is summarized in the flowchart shown in Fig.~\ref{fig:flowchart}. Initially, the computational domain, grids, and computational particles representing real gas particles are generated. At each timestep, ballistic motion and inter-particle collisions are decoupled and processed sequentially. If particle trajectories intersect a surface boundary during the ballistic motion, the gas-surface scattering model is applied to update particle velocities and trajectories. After ballistic motion, inter-particle collisions are computed within each grid cell using the No-Time Counter algorithm, which updates particle velocities while keeping positions unchanged~\cite{bird1994molecular}. Once the flow reaches steady state, macroscopic flow properties are obtained by statistically sampling particle positions and velocities. These samples approximate the velocity distribution function. Macroscopic properties are computed as moments of the velocity distribution function, with density, velocity, and temperature corresponding to the zeroth, first, and second moments, respectively.
%
The corrugated CLL model is applied at surface-boundary intersections to update particle velocities, following the three-step procedure illustrated in Fig.~\ref{fig:flowchart}. 

\noindent\textbf{(1) Sampling the local surface normal}\\
The polar and azimuthal angles $\alpha$ and $\beta$ of the local surface normal, defined by Eq.~\eqref{eq:tiltangle}, are iteratively sampled using the Metropolis-Hastings algorithm~\cite{chib1995understanding}. The Metropolis-Hastings algorithm generates samples iteratively. The initial values are set to $\alpha_0=0$ and $\beta_0=0$. At each iteration, candidate values $\alpha_{\mathrm{new}}=\alpha+\Delta\alpha$ and $\beta_{\mathrm{new}}=\beta+\Delta\beta$ are drawn from Gaussian proposal distributions, clipped to interval $[0,\pi/2)$ for $\alpha$, and adjusted modulo $2\pi$ for $\beta$. The candidate values are accepted with probability:
\begin{equation}
r=\min\left[1,\frac{P_A(\theta_i,0,\alpha_{\mathrm{new}},\beta_{\mathrm{new}},A)}{P_A(\theta_i,0,\alpha,\beta,A)}\right],
\end{equation}
where $P_A$ is given by Eq.~\eqref{eq:tiltangle}. To reduce dependence on the initial values, the first 30 iterations are discarded to incorporate random-walk behavior into the sampling procedure~\cite{pfeiffer2018particle}.

\noindent\textbf{(2) Sampling velocity during local collisions using the CLL model}\\
Once $\alpha$ and $\beta$ are sampled, the incident velocity is decomposed into normal and tangential components, which are then scattered following the conventional CLL model. The reflected normal component $v_{n,r}$ is sampled as
\begin{equation}
v_{n,r} = v_{n,i}\sqrt{(1 - \sigma_n) - \sigma_n\frac{v_{th}^2}{v_{n,i}^2}\ln x_1 + 2\frac{v_{th}}{v_{n,i}}\sqrt{-\sigma_n(1-\sigma_n)\ln x_1}\cos(2\pi x_2)}
\label{eq:cll_norm}
\end{equation}
and the reflected tangential components are given by
\begin{equation}
    v_{t1,r} = v_{t,i}\left(\sqrt{1-\sigma_t} + \frac{v_{th}}{v_{t,i}}\sqrt{-\sigma_t\ln x_3}\cos(2\pi x_4)\right),
\label{eq:cll_tan1}
\end{equation}
\begin{equation}
    v_{t2,r} = \sqrt{-\sigma_t\ln x_5}\, v_{th}\cos(2\pi x_6).
\label{eq:cll_tan2}
\end{equation}
Here, $v_{t1,r}$ represents the in-plane component of the tangential velocity within the plane defined by the incident velocity and the surface normal, while $v_{t2,r}$ represents the out-of-plane component. $\sigma_n$ and $\sigma_t$ are the local normal and tangential accommodation coefficients, $v_{th} = \sqrt{k_b T_w / m_g}$ is the thermal velocity of the surface at wall temperature $T_w$, and $x_1$ to $x_6$ are independent random numbers in $[0,1]$~\cite{padilla2009assessment}.

\noindent\textbf{(3) Verifying the sampled velocity}\\
The sampled velocity is subject to a two-step verification process to determine whether re-collision occurs, in which case the procedure returns to the local surface normal sampling step. First, the velocity direction is evaluated. If it is directed inward toward the surface, re-collision is immediately performed. Second, if the velocity is directed outward, re-collision occurs with the probability defined in Eq.~\eqref{eq:reciprocity}. Otherwise, the particle velocity is updated using the sampled velocity and the particle continues ballistic motion for the remainder of the time step.

\section{Parametric Characteristics of the Corrugated CLL Model}\label{sec:4}

The molecular beam scattering experiments conducted by Rettner et al. are used to validate the corrugated CLL model in reproducing scattering behavior across thermal to structure scattering regimes~\cite{rettner1991angular}. In these experiments, xenon beams are directed on a platinum surface maintained at $T_w=800$~K. Two pairs of angular and energy distributions are provided, corresponding to incident energies $E_i$=1.17~eV and 6.8~eV, both at an incidence angle of $\theta_i=30^\circ$. These cases are reproduced using the conventional and corrugated CLL models. The parameters $\sigma_n$ and $\sigma_t$ of the conventional CLL model are first calibrated to reproduce the angular and energy distributions for $E_i$=1.17~eV. The same parameters are applied for the $E_i$=6.8~eV case, and discrepancies in angular and energy distributions with the experimental results are discussed. The corrugated CLL model is then applied using the same $\sigma_n$ and $\sigma_t$, with the addition of a corrugation factor $A$. The resulting angular and energy distributions for both $E_i$=1.17~eV and 6.8~eV are analyzed.

\begin{figure}[htb]
    \centering
    \begin{subfigure}[b]{0.48\linewidth}
        \centering
        \includegraphics[width=\linewidth]{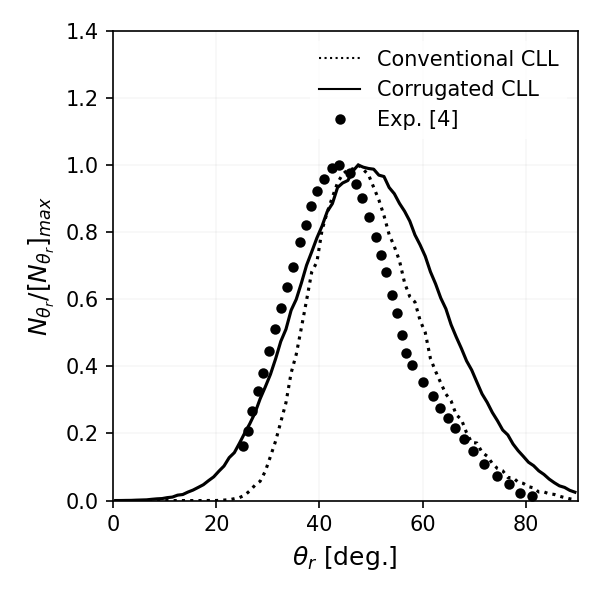}
        \caption{Angular distribution}
        \label{fig:le_intensity}
    \end{subfigure}
    \hfill
    \begin{subfigure}[b]{0.48\linewidth}
        \centering
        \includegraphics[width=\linewidth]{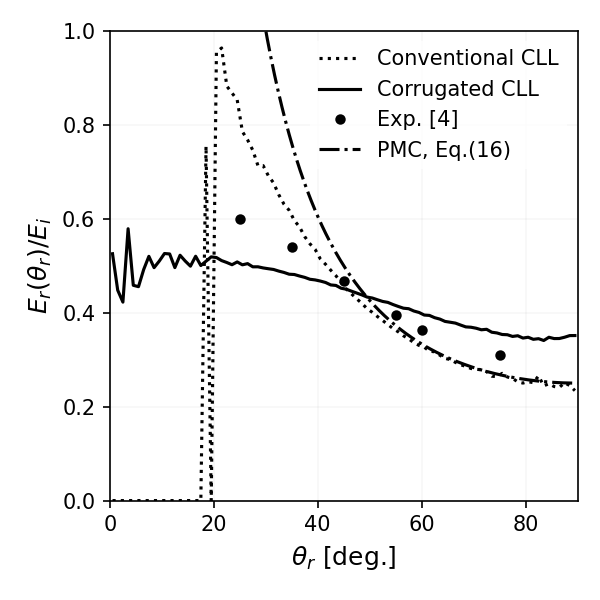}
        \caption{Energy distribution}
        \label{fig:le_energy}
    \end{subfigure}
    \caption{Reflected distributions at $E_i$ = 1.17~eV and $\theta_i$ = 30$^\circ$.}
    \label{fig:le}
\end{figure}

\begin{figure}[htb]
    \centering
    \begin{subfigure}[b]{0.48\linewidth}
        \centering
        \includegraphics[width=\linewidth]{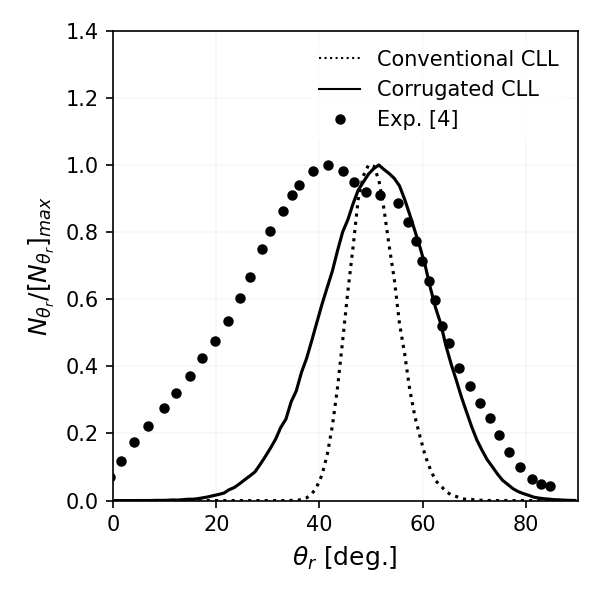}
        \caption{Angular distribution}
        \label{fig:he_intensity}
    \end{subfigure}
    \hfill
    \begin{subfigure}[b]{0.48\linewidth}
        \centering
        \includegraphics[width=\linewidth]{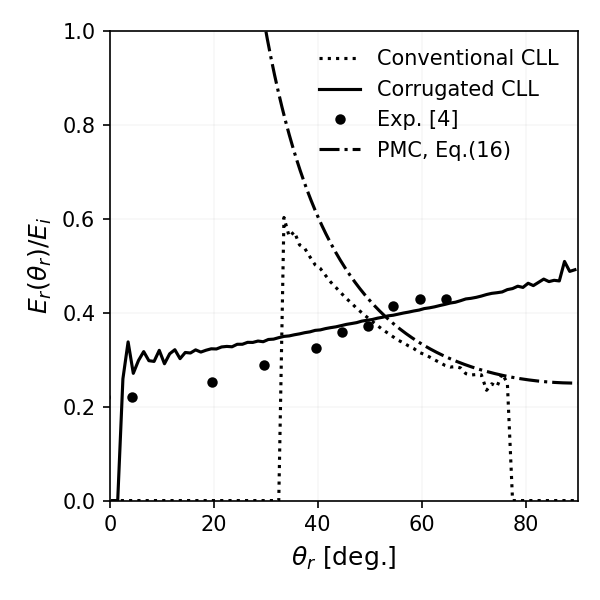}
        \caption{Energy distribution}
        \label{fig:he_energy}
    \end{subfigure}
    \caption{Reflected distributions at $E_i$ = 6.8~eV and $\theta_i$ = 30$^\circ$.}
    \label{fig:he}
\end{figure}


Figure~\ref{fig:le} shows the reflected angular and energy distributions at $E_i$ = 1.17~eV and $\theta_i$ = 30$^\circ$. Experimental results are represented by circles~\cite{rettner1991angular}. In Fig.~\ref{fig:le_intensity}, the angular distribution is defined as the number of reflected particles at each $\theta_r$, denoted as $N_{\theta_r}$, normalized by its maximum value $[N_{\theta_r}]_\text{max}$. The experimental result shows a peak at 43.9$^\circ$, exceeding the specular angle of 30$^\circ$, indicating superspecular reflection. This suggests that the reflected velocity vector is more inclined toward the direction parallel to the local surface than in specular reflection. In Fig.~\ref{fig:le_energy}, the energy distribution is defined as the reflected energy $E_r(\theta_r)$ at each $\theta_r$, normalized by $E_i$. The experimental result shows that $E_r(\theta_r)$ decreases as $\theta_r$ increases. The angular and energy distributions at $E_i=1.17$~eV correspond to thermal scattering regime governed by surface thermal motion~\cite{rettner1991angular}. The low incident energy limits the interaction depth, resulting in dominant normal momentum exchange, while the particle's tangential momentum is nearly conserved due to weak parallel forces. If complete conservation of the particle's tangential momentum is assumed, in-plane reflected particles with $v_{t2,r} = 0$ satisfy $v_{t,i} = v_{t1,r}$. Since $v_{t1,r}$ remains unchanged, $\theta_r$ is only dependent on $v_{n,r}$ as given by 
\begin{equation}
    \tan{\theta_r} = \frac{ v_{t1,r}} { v_{n,r}}.
    \label{eq:theta_r}
\end{equation}
The surface thermal energy corresponding to $T_w$ = 800~K is $E_w$ = 0.07~eV, which is lower than the $E_i$ = 1.17~eV. Energy transfers from the gas to the surface, reducing $v_{n,r}$ relative to $v_{n,i}$ and resulting in a superspecular peak.
The reflected energy $E_r(\theta_r)$ is given by
\begin{equation}
    E_r(\theta_r) =\frac{1}{2} m_g (v_{n,r}^2 + v_{t1,r}^2).
    \label{eq:E_r}
\end{equation} 
Using Eqs.~\eqref{eq:theta_r} and \eqref{eq:E_r}, the relation between $\theta_r$ and $E_r(\theta_r)$ under tangential momentum conservation is obtained as
\begin{equation}
    \frac{E_r(\theta_r)}{E_i} = \left( \frac{ \sin{\theta_i}}{\sin{\theta_r}} \right) ^2,
    \label{eq:pmc}
\end{equation}
commonly referred to as parallel momentum conservation (PMC) relation~\cite{rettner1991angular}. This is shown as the the dash-dotted line in Fig.~\ref{fig:le_energy}. PMC relation accounts for the decrease in $E_r$ as $\theta_r$ increases in the experiment. Yet, the experiment deviates from the PMC relation, indicating imperfect tangential momentum conservation. The results of the conventional CLL model with $\sigma_n=0.8$ and $\sigma_t=0.1$, fitted to experimental data at $E_i$ = 1.17~eV, are shown as dotted lines in Fig.~\ref{fig:le}. The $\sigma_n$ is significantly higher than $\sigma_t$, indicating dominant normal energy exchange consistent with the PMC relation. The conventional CLL model exhibits a superspecular peak at $\theta_r = 47.5^\circ$, deviating by approximately $3.6^\circ$ from the experimental peak. The energy distribution from conventional CLL model is closer to the PMC relation than the experimental result, suggesting that the experimental result corresponds to a higher $\sigma_t$ than the value of 0.1. Nevertheless, the conventional CLL model captures the negative slope of $E_r$ with increasing $\theta_r$, characteristic of thermal scattering.

\begin{figure}
    \centering
    \includegraphics[width=0.8\linewidth]{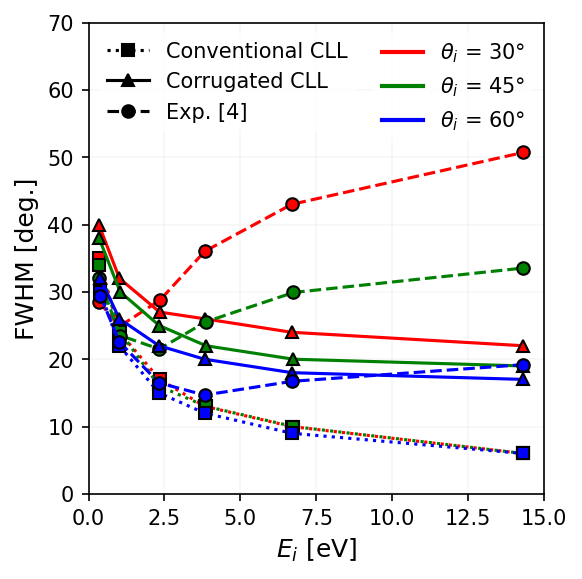}
    \caption{FWHM as a function of $E_i$ at $\theta_i=$30$^\circ$, 45$^\circ$, and 60$^\circ$.}
    \label{fig:fwhm}
\end{figure}
\begin{figure}
    \centering
    \includegraphics[width=0.8\linewidth]{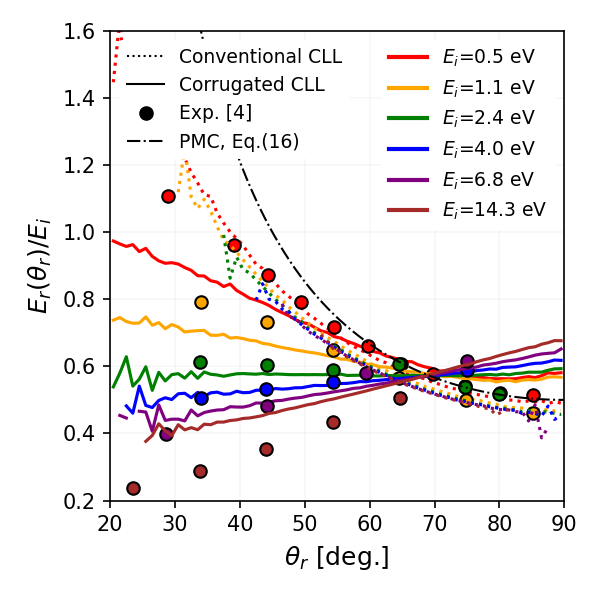}
    \caption{$E_r$ as a function of $\theta_r$ for $E_i=$0.5~eV to 14.3~eV, $\theta_i=45^\circ$.}
    \label{fig:energy_dist}
\end{figure}

The angular and energy distributions at the increased $E_i$=6.8~eV and $\theta_i=30^\circ$ are presented in Fig.~\ref{fig:he}. The angular distribution in Fig.~\ref{fig:he_intensity} shows that the experimental result, shown by circles, peaks at a superspecular angle of $41.6^\circ$. It exhibits a bimodality with a secondary peak around $60^\circ$, indicative of rainbow scattering observed in structure scattering regimes~\cite{kleyn1991rainbow,rettner1991angular}. This bimodal behavior leads to a broader angular width at $E_i$=6.8~eV compared to the $E_i$=1.17~eV case, which is quantified by the full-width half-maximum (FWHM) of the angular distribution. Figure~\ref{fig:fwhm} shows the dependence of FWHM on $E_i$, evaluated at 0.5, 1.17, 2.4, 4.0, 6.8, and 14.3~eV, for $\theta_i=30^\circ$, $45^\circ$, and $60^\circ$. For the experimental results at $\theta_i=30^\circ$, indicated by red circles, the FWHM decreases from $E_i$=0.5~eV to 1.17~eV, then increases for $E_i$>1.17~eV. This nonlinear behavior demonstrates a transition from thermal to structure scattering. For $E_i<1.17$~eV, the initial narrowing of the angular distribution is characteristic of thermal scattering. In this regime, increasing $E_i$ reduces the influence of surface thermal motion, leading to a decrease in angular width. At higher $E_i$>1.17~eV, increasing $E_i$ enhances surface penetration and amplifies the effect of surface corrugation. This leads to broader angular distributions indicative of structure scattering~\cite{livadiotti2020review,oman1968numerical}. The angular distribution at $E_i$=6.8~eV, predicted by the conventional CLL model with $\sigma_n=0.8$, $\sigma_t=0.1$, is shown as a dotted line in Fig.~\ref{fig:he_intensity}. It predicts a superspecular angle of 49.5$^\circ$ but a narrower distribution than at $E_i$=1.17~eV. The conventional CLL model yields monotonically decreasing FWHM with increasing $E_i$ for all $\theta_i$, as shown in Fig.~\ref{fig:fwhm} with squares. It agrees with the experiments at low $E_i$<1.17~eV, but fails to reproduce the increase in FWHM observed at higher $E_i$>1.17~eV. This is because the conventional CLL model, as defined in Eqs.~\eqref{eq:cll}, accounts only for surface temperature and does not consider surface corrugation.
Figure~\ref{fig:he_energy} illustrates the reflected energy distribution at $E_i$=6.8~eV and $\theta_i=30^\circ$. The experimental result, shown by circles, exhibit increasing $E_r$ with $\theta_r$ and deviates from the PMC relation. This deviation indicates a breakdown of tangential momentum conservation, as surface corrugation induces tangential forces that violate the assumptions of the PMC relation~\cite{goodman1967three}. In contrast, the conventional CLL model with fixed $\sigma_n=0.8$, $\sigma_t=0.1$, shown by a dotted line, retains the negative slope consistent with the PMC relation. Figure~\ref{fig:energy_dist} shows how the energy distribution evolves with increasing $E_i$ from 0.5~eV to 14.3~eV. The incident angle $\theta_i=45^\circ$ is used, as the experimental energy distributions over a range of $E_i$ are available only at this angle in Rettner et al.~\cite{rettner1991angular}. The experimental result follows the PMC relation at the low energy of $E_i$=0.5~eV. The slope gradually increases with $E_i$ and becomes positive beyond $E_i$=4.0~eV. The change in the energy distribution slope from negative to positive reflects increasing deviation from tangential momentum conservation, indicating a transition from thermal to structure scattering. The conventional CLL model shown with dotted lines, however, maintains a negative slope regardless of $E_i$, failing to capture this transition.


The corrugated CLL model is evaluated using a fixed corrugation factor $A = 0.1$ to reproduce the experimental results at $E_i = 1.17$~eV and 6.8~eV. At $E_i=1.17$~eV in Fig.~\ref{fig:le_intensity}, the corrugated CLL model shown in solid line produces a superspecular peak at $\theta_r = 47.5^\circ$. It exhibits a broader angular distribution than the conventional CLL model due to randomization of reflected directions by surface corrugation. In Fig.~\ref{fig:he_intensity}, the corrugated CLL model retains the superspecular peak at $\theta_r=51.5^\circ$. It predicts a wider angular width than the conventional CLL model, though narrower than the experiment. As shown in Fig.~\ref{fig:fwhm} with triangles, the FWHM predicted by the corrugated CLL model decreases monotonically with $E_i$. While the conventional CLL model predicts that the FWHM decreases toward zero, the corrugated CLL model predicts convergence to approximately $20^\circ$ due to enhanced angular scattering in the structure scattering regime. Although the corrugated CLL model incorporates surface corrugation, it still fails to reproduce the increase in FWHM observed at $E_i > 1.17$~eV, similar to the conventional model. This is because the corrugated CLL model applies a fixed corrugation factor $A$. In practice, higher $E_i$ increases the penetration depth, thereby enhancing the effective surface corrugation. In Figs.~\ref{fig:le_energy} and \ref{fig:he_energy}, the energy distributions predicted by the corrugated CLL model are shown as solid lines. At $E_i = 1.17$~eV, the corrugated CLL model shows a flatter slope than the experimental result, but the slope remains negative. At $E_i=6.8$~eV, the corrugated CLL model reproduces the experimentally observed positive slope. The change in slope with increased $E_i$ is attributed to the effect of local surface orientation. When the local surface tilts toward the incident beam, the normal velocity component increases, leading to greater energy loss; such orientations also result in scattering toward lower angles. When the surface tilts away from the incident direction, energy loss and scattering angles both increase. The solid lines in Fig.~\ref{fig:energy_dist} show that the corrugated CLL model captures the gradual change in slope from negative to positive with increasing $E_i$. This demonstrates that the slope change in the energy distribution across the thermal-to-structure scattering transition can be attributed to the surface corrugation.

The energy accommodation coefficients are calculated by combining the angular and energy distributions, under the assumption that the incident beam and the scattered particles at each $\theta_r$ have uniform velocities. Due to the absence of thermal velocity in the incident beam, two definitions of the energy accommodation coefficient are introduced. The first definition employs total velocities of incident and reflected particles, as given in Eq.~\eqref{eq:eac}, and is referred to as the total energy accommodation coefficient~\cite{goodman1974thermal}. It quantifies the kinetic energy lost by the incident beam during collision, which corresponds to the energy transferred to the surface. The second definition subtracts the mean velocity component to extract the thermal velocity, expressed as:
\begin{align} 
\sigma_n &= \frac{0 - \langle \frac{1}{2} m_g (v_{r,n}-\langle v_{r,n} \rangle )^2 \rangle}{0 - E_{w,n}}, \label{eq:neac} \\ 
\sigma_t &= \frac{0 - \langle \frac{1}{2} m_g (v_{r,t}-\langle v_{r,t} \rangle )^2 \rangle}{0 - E_{w,t}}, \label{eq:teac}
\end{align}
since the thermal energy in the incident beam is negligible~\cite{liang2018physical}. This definition characterizes the increase in thermal velocity component among scattered particles and is referred to as the normal and tangential thermal energy accommodation coefficient.
\begin{figure}
    \centering
    \includegraphics[width=0.8\linewidth]{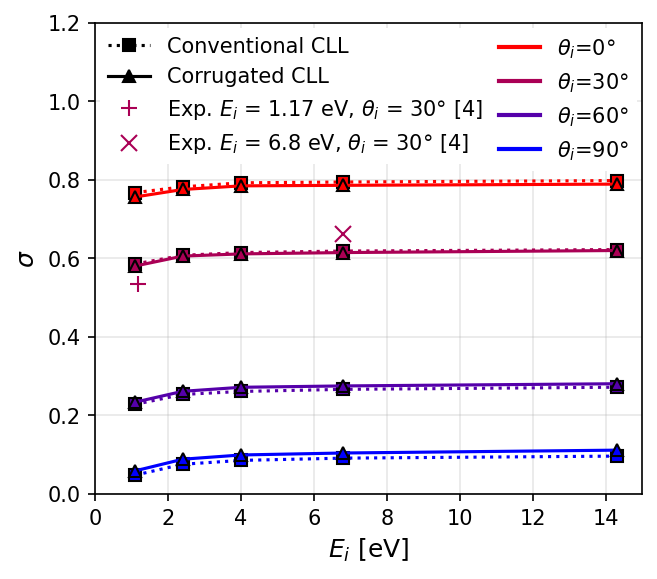}
    \caption{Total energy accommodation coefficient $\sigma$, calculated using Eq.~\eqref{eq:eac}, for $E_i$=1.17~eV to 14.3~eV and $\theta_i=0^\circ$, $30^\circ$, $60^\circ$, and $90^\circ$.}
    \label{fig:eac}
\end{figure}
Figure~\ref{fig:eac} illustrates the total energy accommodation coefficient $\sigma$ for $E_i=$ 1.17~eV to 14.3~eV at $\theta_i=0^\circ$, $30^\circ$, $60^\circ$, and $90^\circ$. The case of $E_i=0.5$~eV is excluded because the denominator in Eq.~\eqref{eq:eac} approaches zero, resulting in a singularity. The dotted lines represent the conventional CLL model, and the solid lines represent the corrugated CLL model. Experimental results that provide both angular and energy distributions are available only at $E_i$=1.17~eV and 6.8~eV with $\theta_i=30^\circ$, as shown in Figs.~\ref{fig:le} and \ref{fig:he}~\cite{rettner1991angular}. The corresponding values of $\sigma$ are plotted using plus and cross symbols. Overall, the conventional and corrugated CLL models yield nearly identical $\sigma$ across all $\theta_i$, and both show good agreement with the experimental results at $\theta_i=30^\circ$. This indicates that the corrugation factor of $A=0.1$ exerts limited influence on the kinetic energy loss of gas particles during scattering. However, small differences are observed depending on $\theta_i$. At lower angles of incidence, such as $\theta_i=0^\circ$ and 30$^\circ$, the corrugated CLL yields slightly lower $\sigma$ than the conventional CLL model. This is attributed to surface corrugation redirecting velocities initially normal to the surface into tangential directions. In contrast, at higher angles of incidence, surface corrugation promotes energy exchange by increasing the effective surface-normal component, resulting in slightly higher $\sigma$ in the corrugated CLL model compared to the conventional CLL model.
\begin{figure}[htbp]
    \centering
    \begin{subfigure}[b]{0.48\linewidth}
        \centering
        \includegraphics[width=\linewidth]{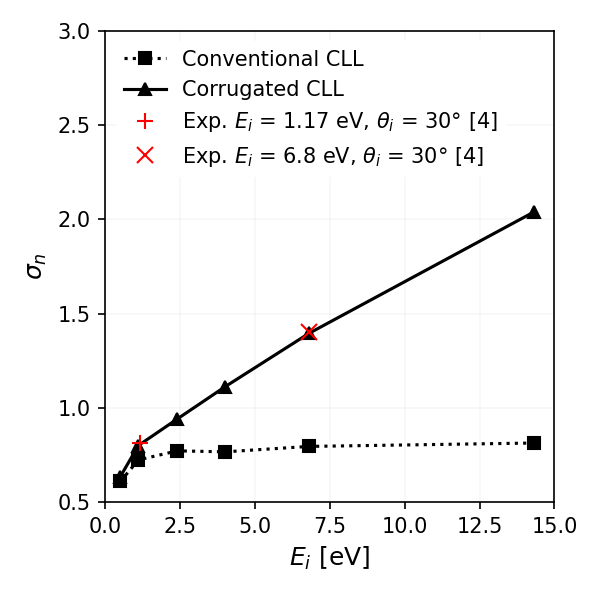}
        \caption{Normal thermal energy accommodation coefficient $\sigma_n$}
        \label{fig:nacc}
    \end{subfigure}
    \hfill
    \begin{subfigure}[b]{0.48\linewidth}
        \centering
        \includegraphics[width=\linewidth]{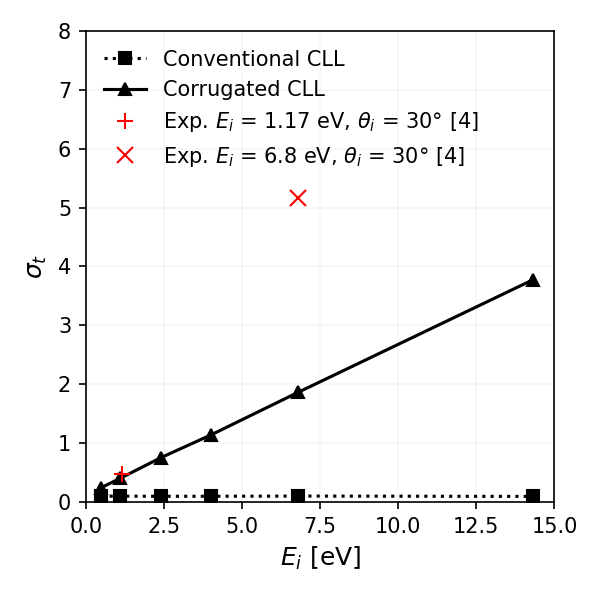}
        \caption{Tangential thermal energy accommodation coefficient $\sigma_t$}
        \label{fig:tacc}
    \end{subfigure}
    \caption{Thermal energy accommodation coefficients $\sigma_n$ and $\sigma_t$ calculated using Eqs.~\eqref{eq:neac} and \eqref{eq:teac} for $E_i$=0.5~eV to 14.3~eV and $\theta_i=30^\circ$}
    \label{fig:ntacc}
\end{figure}
Figure~\ref{fig:ntacc} presents the thermal accommodation coefficients $\sigma_n$ and $\sigma_t$ from Eqs.~\eqref{eq:neac} and \eqref{eq:teac} for $E_i=$0.5~eV to 14.3~eV at $\theta_i$=30$^\circ$. The experimentally obtained $\sigma_n$ and $\sigma_t$ at $E_i=1.17$~eV and 6.8~eV are also presented in cross and plus symbols. Both $\sigma_n$ and $\sigma_t$ are higher at $E_i=$6.8~eV compared to $E_i=$1.17~eV, indicating an increased thermal velocity component in the reflected particles under the structure scattering regime. The conventional CLL model fails to reproduce increasing $\sigma_n$ and $\sigma_t$ with $E_i$, as both values asymptotically approach the input values of 0.8 and 0.1. Importantly, the conventional CLL model cannot reproduce values of $\sigma_n$ and $\sigma_t$ exceeding unity, as such values are not permitted in Eqs.~\eqref{eq:cll_norm}-\eqref{eq:cll_tan2}. This is because the thermal velocity of reflected particles in the conventional CLL model depends only on the surface temperature. In contrast, the corrugated CLL model captures the increase in $\sigma_n$ and $\sigma_t$ with increasing $E_i$ and reproduces values exceeding unity. This is because surface corrugation broadens the reflected directions, increasing the thermal component in the reflected velocity. The accommodation coefficients increase nearly linearly with $E_i$, which is consistent with the linear scaling of the molecular thermal energy in Eqs.~\eqref{eq:neac} and \eqref{eq:teac} under fixed surface corrugation. While the corrugated CLL model agrees well with the experimental $\sigma_n$, it underpredicts $\sigma_t$, particularly at high $E_i$. This discrepancy is attributed to the use of a fixed corrugation factor $A$, which does not account for the increase in effective surface corrugation resulting from deeper particle penetration at higher energies. In addition, the random Gaussian surface geometry used in the model assumes spatially homogeneous roughness, whereas real surfaces often exhibit lattice structures with position-dependent local slopes that are not captured in the current formulation.


\section{Effect of Surface Corrugation on the Macroscopic Flow Field}\label{sec:5}

\subsection{External Flow Around a Cylinder} \label{sec:5_1}

To investigate the effect of surface corrugation on drag in VLEO conditions, a DSMC simulation is performed with the conventional and corrugated CLL models applied to the cylinder surface. The freestream number density and temperature, corresponding to a mixture of atomic oxygen and nitrogen under rarefied conditions at a VLEO altitude of 200 km, are obtained from the NRLMSISE-2.0 model and summarized in Table~\ref{tab:freestream}~\cite{emmert2021nrlmsis}. However, experimental data for atomic oxygen and nitrogen scattering at orbital velocities, required for modeling of gas–surface interactions in VLEO, are currently unavailable. Therefore, the freestream gas and surface material are assumed to be xenon and platinum, as used in the experiment by Rettner et al.~\cite{rettner1991angular}. The freestream velocity of 3160~m/s is assumed, which corresponds to an incident energy of $E_i = 6.8$~eV. Three scattering models with increasing degrees of surface corrugation are considered. Case A applies the conventional CLL model with $\sigma_n=0.8$ and $\sigma_t=0.1$, while Case B uses the corrugated CLL model with $A=0.1$, $\sigma_n=0.8$, and $\sigma_t=0.1$, as in Section~\ref{sec:4}. Additionally, Case C increases the corrugation factor to $A=0.2$ to examine the effect of increased surface corrugation. The scattering models for each case are summarized in Table~\ref{tab:scattering_models}. Intermolecular collisions are modeled using the variable hard sphere (VHS) parameters for xenon, listed in Table~\ref{tab:xenon_vhs}. The cylinder diameter is 0.1~m, and the simulation employs an average of 65 million particles over 4 million grid cells, with a time step of $10^{-7}$~s to ensure accurate resolution of collisions.

\begin{table}[htbp]
\centering
\caption{Freestream condition}
\label{tab:freestream}
\begin{tabular}{lc}
    \hline
    Parameter & Value \\
\hline
Number density          & $6.7 \times 10^{15}$ m$^{-3}$ \\
Velocity     & 3160~m/s               \\
Temperature             & 800~K                  \\ \hline
\end{tabular}
\end{table}

\begin{table}[htb]
  \centering
  \caption{Xenon VHS Parameters}
  \label{tab:xenon_vhs}
  \begin{tabular}{lc}
    \hline
    Parameter & Value \\
    \hline
    $d_{\mathrm{ref}}$ & $5.74 \times 10^{-10}\,\mathrm{m}$ \\
    $\omega$ & 0.85 \\
    $T_{\mathrm{ref}}$ & 273.15~K \\
    \hline
  \end{tabular}
\end{table}
     
\begin{table}[htbp]
    \centering
    \caption{Scattering models employed for each case}
    \label{tab:scattering_models}
    \begin{tabular}{cl}
    \hline
    Case & Scattering model \\ \hline
    A & Conventional CLL ($\sigma_n=0.8$, $\sigma_t=0.1$) \\
    B & Corrugated CLL ($A=0.1$, $\sigma_n=0.8$, $\sigma_t=0.1$) \\
    C & Corrugated CLL ($A=0.2$, $\sigma_n=0.8$, $\sigma_t=0.1$) \\ \hline
    \end{tabular}
\end{table}

\clearpage


\begin{figure}[htb]
  \centering
  \begin{subfigure}[b]{0.48\textwidth}
    \includegraphics[width=\linewidth]{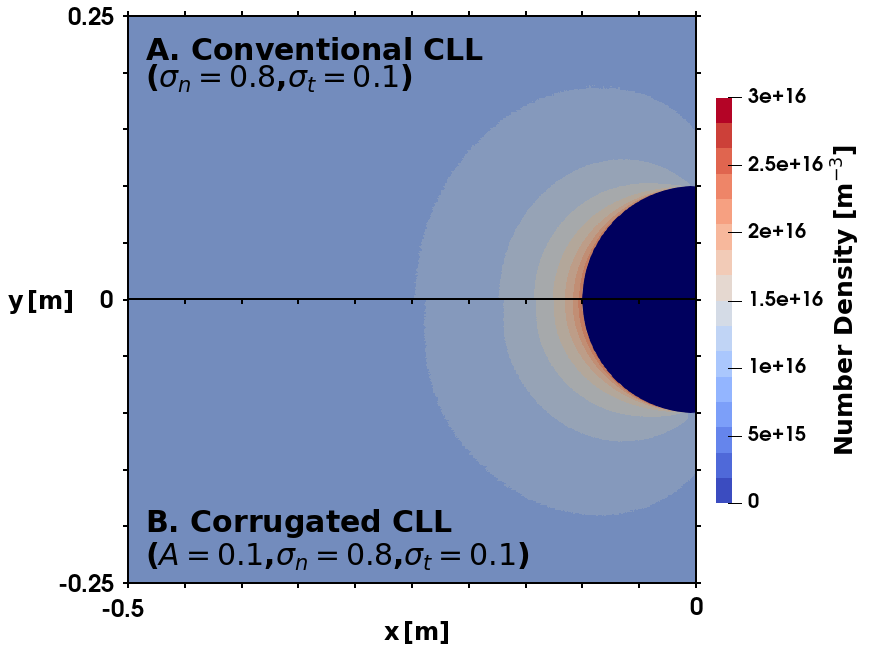}
    \caption{Number density: Cases A-B}
    \label{fig:cyl_n_ab}
  \end{subfigure}
  \hfill
  \begin{subfigure}[b]{0.48\textwidth}
    \includegraphics[width=\linewidth]{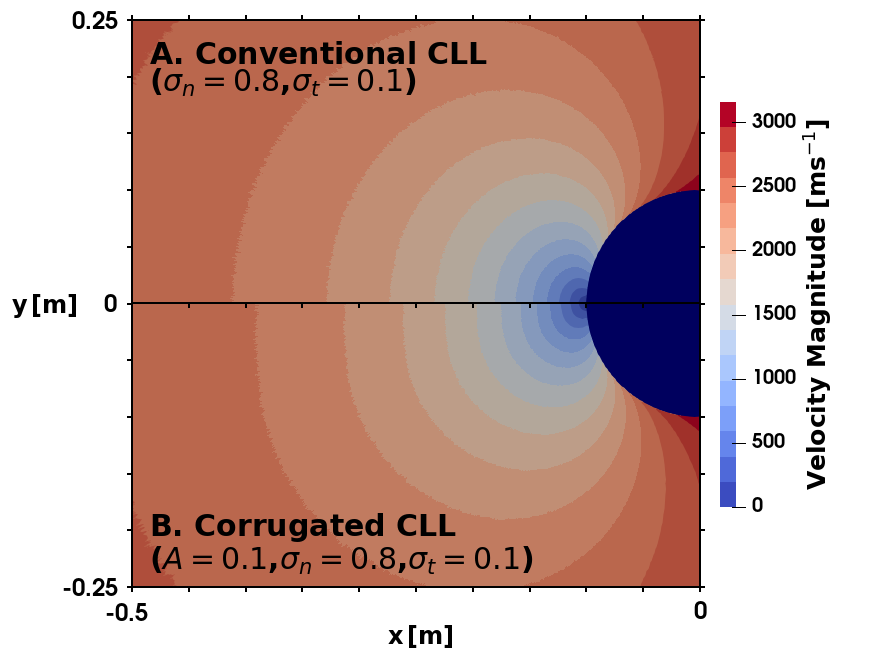}
    \caption{Velocity magnitude: Cases A-B}
    \label{fig:cyl_v_ab}
  \end{subfigure}
  
  \bigskip
  
  \begin{subfigure}[b]{0.48\textwidth}
    \includegraphics[width=\linewidth]{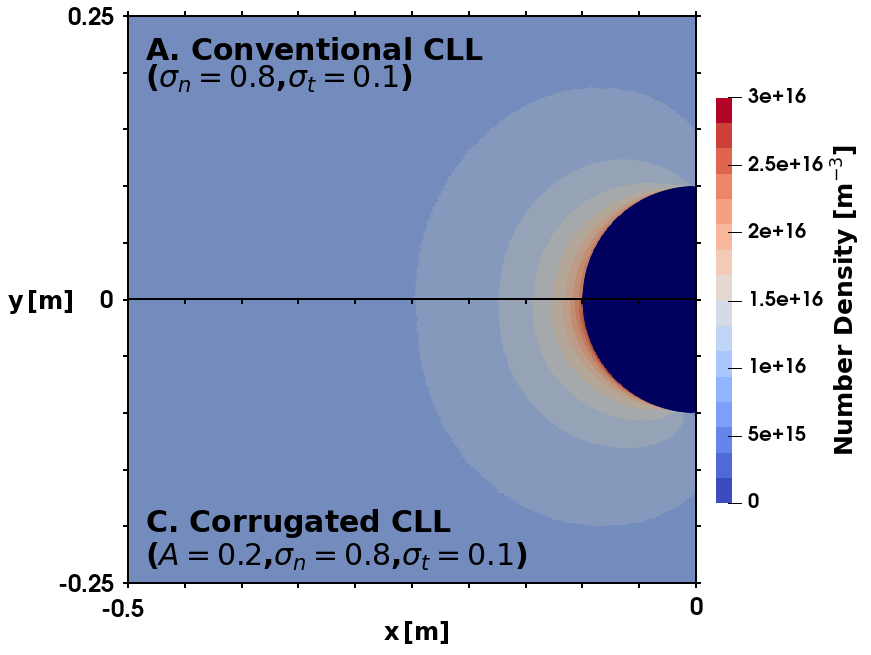}
    \caption{Number density: Cases A-C}
    \label{fig:cyl_n_ac}
  \end{subfigure}
  \hfill
  \begin{subfigure}[b]{0.48\textwidth}
    \includegraphics[width=\linewidth]{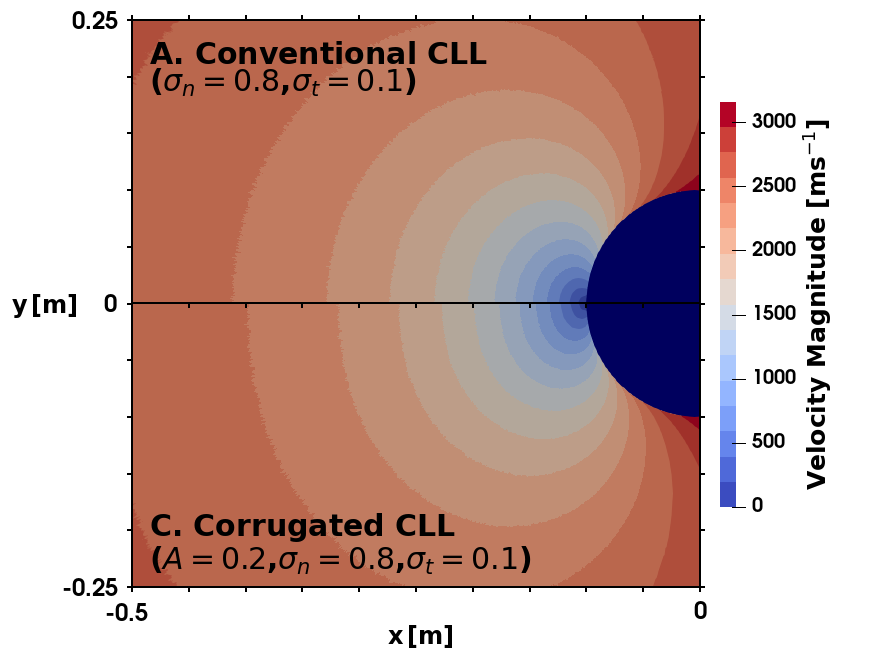}
    \caption{Velocity magnitude: Cases A-C}
    \label{fig:cyl_v_ac}
  \end{subfigure}

  \caption{Flow around a cylinder: number density and velocity magnitude}
  \label{fig:cyl_all}
\end{figure}
\clearpage

\begin{figure}[ht]
  \centering
  \begin{subfigure}[b]{0.48\textwidth}
    \includegraphics[width=\linewidth]{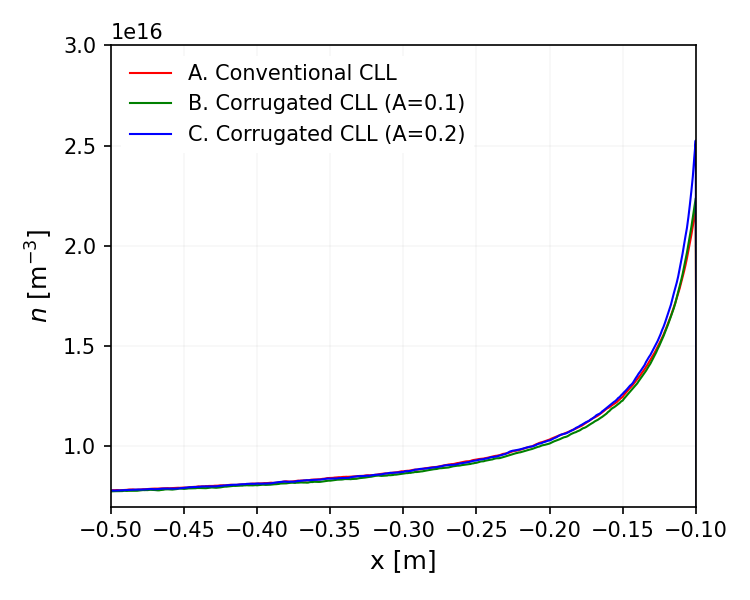}
    \caption{Number density}
    \label{fig:stag_n}
  \end{subfigure}
\hfill
  \begin{subfigure}[b]{0.48\textwidth}
    \includegraphics[width=\linewidth]{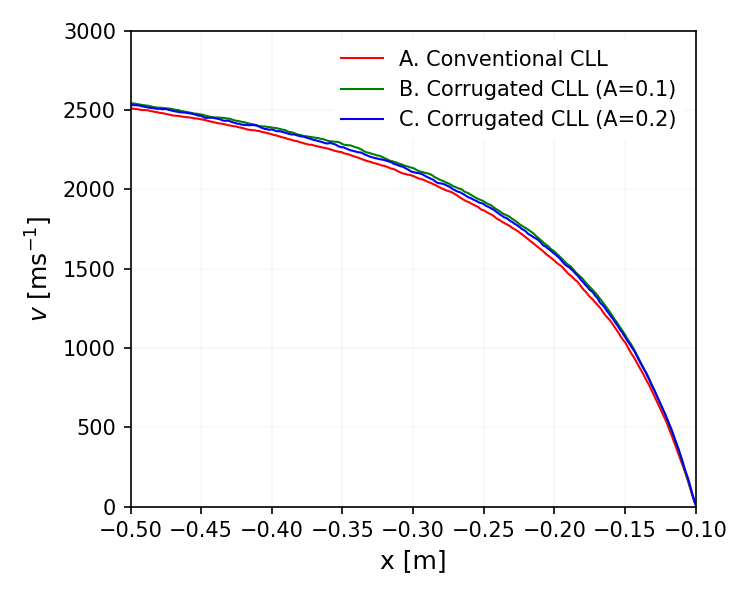}
    \caption{Velocity magnitude}
    \label{fig:stag_v}
  \end{subfigure}

    \caption{Number density and velocity along the stagnation line}
    \label{fig:stag}
\end{figure}

\begin{table}[htbp]
    \centering
    \caption{Number density at the stagnation point}
    \label{tab:n_stag}
    \begin{tabular}{ccccc}
        \hline
        Case A & Case B & Case C \\ 
        \hline
        $2.176 \times 10^{16}$ & $2.237 \times 10^{16}$ & $2.525 \times 10^{16}$  \\
        \hline
    \end{tabular}
\end{table}

\begin{figure}[ht]
  \centering
  \begin{subfigure}[b]{0.45\textwidth}
    \includegraphics[width=\linewidth]{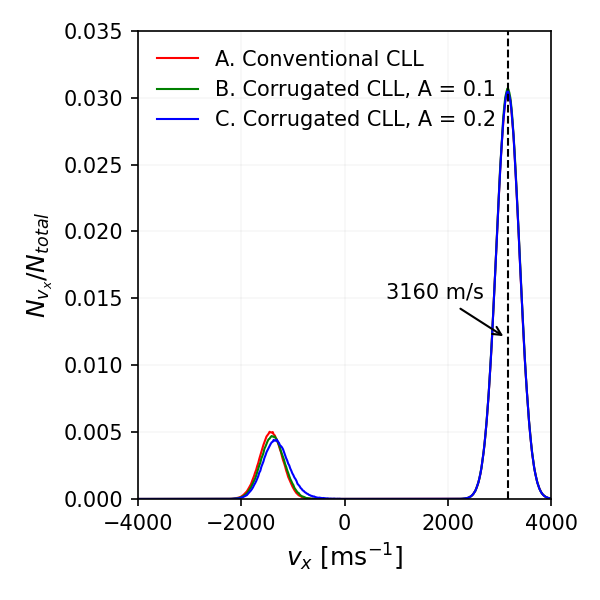}
    \caption{$x$-velocity}
    \label{fig:histo_x}
  \end{subfigure}
  \quad
  \begin{subfigure}[b]{0.45\textwidth}
    \includegraphics[width=\linewidth]{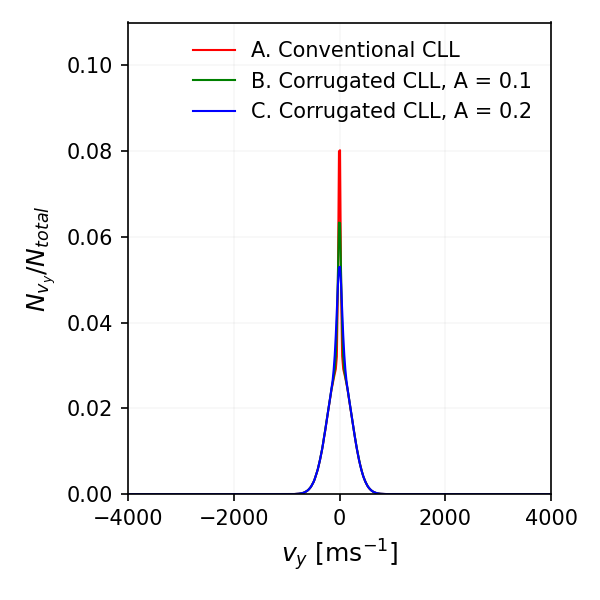}
    \caption{$y$-velocity}
    \label{fig:histo_y}
  \end{subfigure}

    \caption{Velocity distribution function in $x$ and $y$ directions at the stagnation point}
    \label{fig:histo}
\end{figure}

Figures~\ref{fig:cyl_all} and \ref{fig:stag} show the number density and velocity magnitude around the cylinder and along the stagnation line. In all cases, number density increases and velocity decreases near the surface along the stagnation line. As shown in Fig.~\ref{fig:stag_n}, the number density profiles are nearly identical across all three cases. However, near the stagnation point at $x = -0.1$~m, only Case C exhibits a distinct increase in number density. At the stagnation point, Case B shows a 2\% increase compared to Case A, while Case C exhibits a larger increase of 16\%, as summarized in Table~\ref{tab:n_stag}. Figure~\ref{fig:stag_v} shows that the velocity magnitude along the stagnation line is higher in Cases B and C than in Case A, while the difference between Cases B and C is negligible. Velocity distribution functions at the stagnation point are shown in Figure~\ref{fig:histo}. Figure~\ref{fig:histo_x} presents the $x$-velocity distribution, which exhibits a bimodal profile with a freestream peak at $v_x=3160$~m/s and a reflected peak near $-1500$~m/s. As surface corrugation increases from Case A to C, the reflected peak shifts toward zero and decreases in magnitude, indicating a reduction in normal velocity component. Figure~\ref{fig:histo_y} shows a symmetric $y$-velocity distribution about zero. The fraction of particles with $v_y=0$ decreases from Case A to C, and the distribution broadens, indicating an increase in the tangential velocity component induced by surface corrugation. The reduction in the normal velocity and enhancement of the tangential component lead to longer particle residence time near the surface, resulting in higher number density at the stagnation point. Additionally, lower normal velocities reduce the probability of reflection opposite to the freestream direction, thereby increasing the average velocity along the stagnation line.

\begin{figure}[ht]
  \centering
  \begin{subfigure}[b]{0.48\textwidth}
    \includegraphics[width=\linewidth]{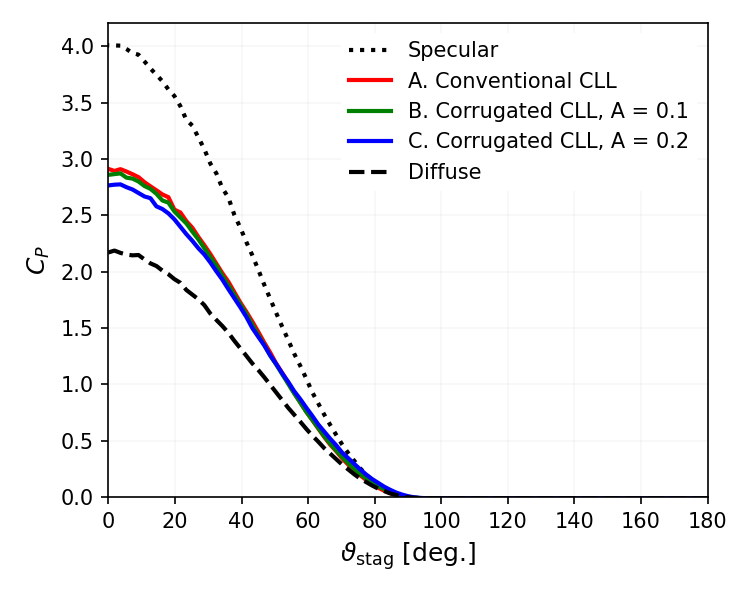}
    \caption{Pressure coefficient}
    \label{fig:press_coeff}
  \end{subfigure}
\hfill
  \begin{subfigure}[b]{0.48\textwidth}
    \includegraphics[width=\linewidth]{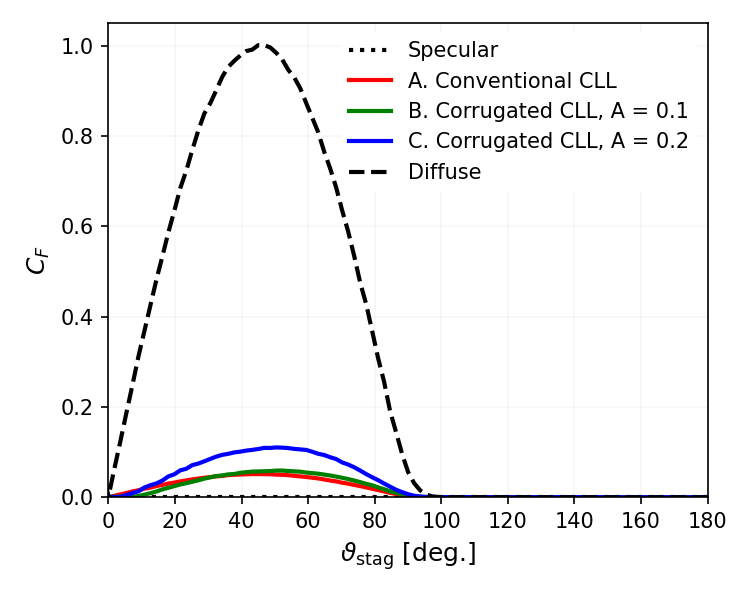}
    \caption{Friction coefficient}
    \label{fig:fric_coeff}
  \end{subfigure}
 \vspace{0.5cm}
  \begin{subfigure}[b]{0.48\textwidth}
    \includegraphics[width=\linewidth]{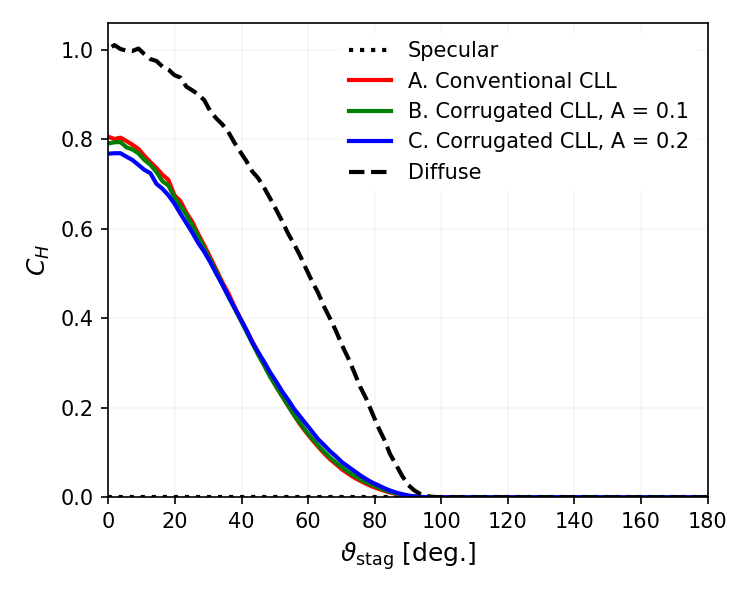}
    \caption{Heat transfer coefficient}
    \label{fig:heat_coeff}
  \end{subfigure}
    \caption{Cylinder surface properties for different scattering models.}
    \label{fig:surf_coeff}
\end{figure}

Figure~\ref{fig:surf_coeff} compares the pressure, friction, and heat transfer coefficients along the cylinder surface as a function of angle $\vartheta_{stag}$ from the stagnation point. These coefficients are defined as $C_P = 2(P-P_\infty)/\rho_\infty U_\infty^2$, $C_F = 2\tau_w /\rho_\infty U_\infty^2$, and $C_H = 2q_w / \rho_\infty U_\infty^3$, where $P$, $\tau_w$, and $q_w$ denote the local pressure, wall shear stress, and wall heat flux, respectively, and $P_\infty$, $\rho_\infty$, and $U_\infty$ represent the freestream pressure, density, and velocity. Solid red, green, and blue lines represent Case A, B, and C using the conventional CLL model and corrugated CLL models with $A=0.1$ and $A=0.2$, respectively. Dotted and dashed lines correspond to specular and diffuse reflections, which represent the two limiting behaviors of the Maxwell model commonly used in DSMC simulations for VLEO applications~\cite{moon2023performance}. As $\vartheta_{stag}$ increases, the projected surface area and the normal component of the incident velocity decrease, reducing the number of impinging particles and the normal momentum and energy transferred to the surface. This leads to a decrease in both $C_P$ and $C_H$. Meanwhile, the tangential velocity component increases with $\vartheta_{stag}$, enhancing momentum transfer along the surface and causing $C_F$ to exhibit a peak between 0$^\circ$ and 90$^\circ$. Specular reflection yields the highest $C_P$ but zero $C_F$ and $C_H$ due to the absence of energy and tangential momentum exchange. In contrast, diffuse reflection minimizes $C_P$ and maximizing $C_F$ and $C_H$. Cases A, B, and C all employ CLL-based models with $\sigma_n = 0.8$ and $\sigma_t = 0.1$, assuming partial accommodation and producing results between the two limiting cases. Case B, with $A=0.1$, shows less than 1\% deviation in all surface coefficients compared to Case A. This indicates that this level of surface corrugation does not significantly affect surface properties under the given accommodation coefficients. In contrast, Case C with $A=0.2$ shows approximately a 5\% reduction in the peak values of both $C_P$ and $C_H$, while the peak $C_F$ nearly doubles relative to Case A. The reduction in $C_P$ results from a decrease in the normal component of the reflected velocity, which leads to reduced normal momentum transfer. The reduction in $C_H$ is associated with the smaller reflected velocity magnitude, governed by the total energy accommodation coefficient $\sigma$. As shown in Fig.~\ref{fig:eac}, $\sigma$ at $\theta_i = 0$ is slightly lower in the corrugated CLL model than in the conventional CLL model, resulting in reduced energy transfer to the surface. In contrast, Fig.~\ref{fig:ntacc} shows that $\sigma_t$ at $E_i$=6.8~eV is significantly higher for the corrugated CLL model, which enhances tangential momentum transfer and thereby increases $C_F$.
\begin{figure}
    \centering
    \includegraphics[width=0.5\linewidth]{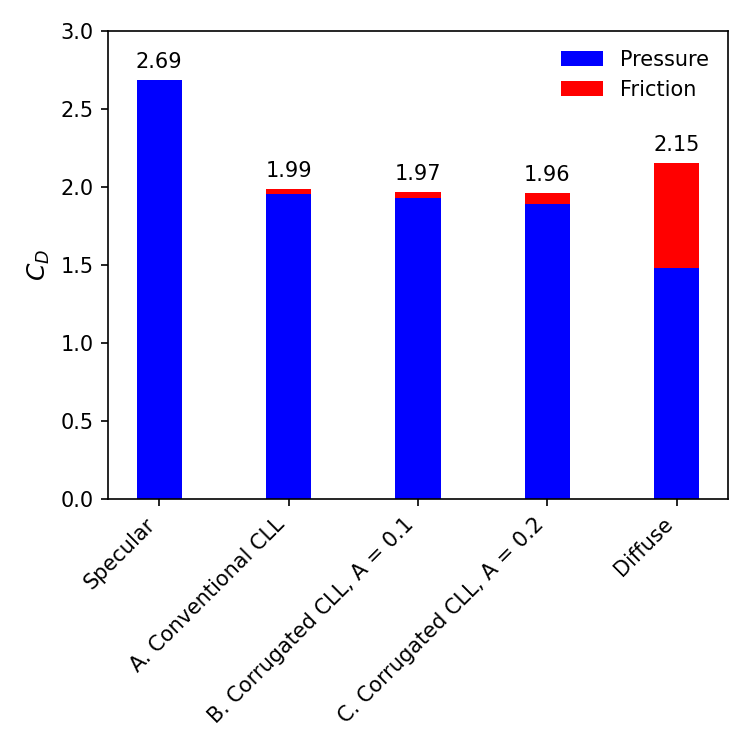}
    \caption{Drag coefficients for different scattering models.}
    \label{fig:dragcoeff}
\end{figure}


Figure~\ref{fig:dragcoeff} presents the drag coefficient $C_D$ computed using different scattering models. The $C_D$ is defined as $C_D=2F_D / \rho_\infty m_g U_\infty^2 L$, where $F_D$ is the total drag force integrated from surface pressure and shear stress, and $L$ is the cylinder diameter. The pressure and friction contributions to $C_D$ are shown as blue and red areas, respectively. Specular reflection yields the highest $C_D$ of 2.69, entirely due to pressure drag. In contrast, diffuse reflection exhibits the lowest pressure drag and the largest friction drag, resulting in a total $C_D$ of 2.15. Cases A to C exhibit lower $C_D$ than both limiting cases. In Cases A to C, partial normal accommodation reduces pressure drag compared to specular reflection, while low tangential accommodation limits friction drag relative to diffuse reflection. Among Cases A to C, total $C_D$ values are similar. However, as surface corrugation increases, the pressure contribution decreases while the friction contribution increases. The friction contribution rises from 1.73\% in Case A to 2.01\% in Case B and 3.77\% in Case C. The surface corrugation used in this analysis remain small to minimize trapping-desorption effects, resulting in minimal variation in total $C_D$. Nevertheless, further increases in surface corrugation are expected to reduce pressure drag and enhance friction, potentially altering the total $C_D$. These results highlight the importance of accounting for surface corrugation in accurate drag prediction for VLEO applications.

\subsection{Internal Flow in Ducted Intake} \label{sec:5_2}

\begin{figure}
    \centering
    \includegraphics[width=0.8\linewidth]{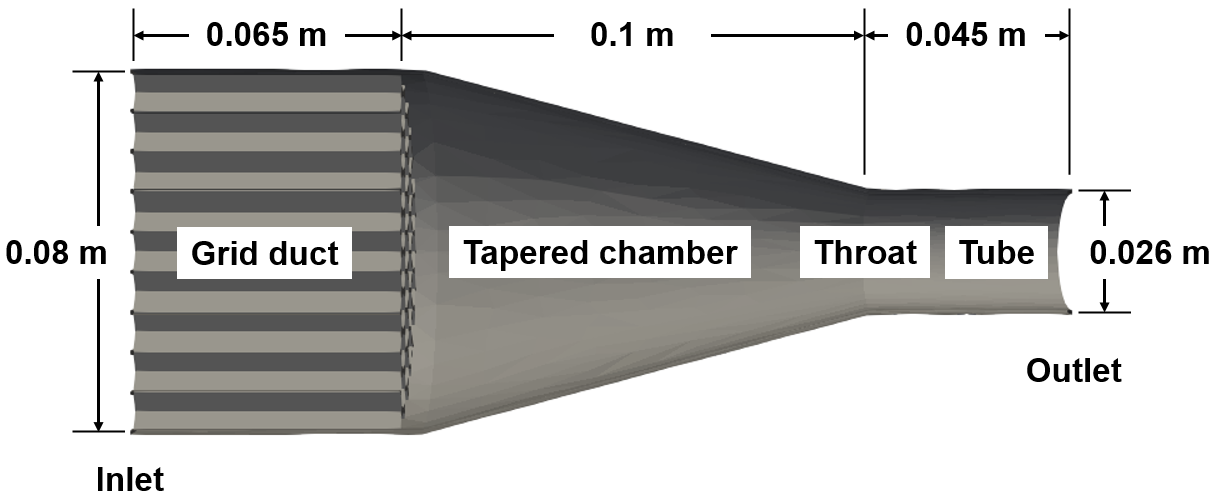}
    \caption{ABEP intake geometry.}
    \label{fig:intake_geometry}
\end{figure}
Atmosphere-breathing electric propulsion (ABEP) systems generate thrust using captured atmospheric gas, enabling drag compensation for satellites operating in VLEO~\cite{nishiyama2003air,moon2024design,moon2025operational}. The feasibility of such systems depends on the intake performance, typically characterized by the capture efficiency $\eta = \frac{N_{\text{throat}}}{N_{\text{inlet}}}$ and compression ratio $CR = \frac{n_{\text{throat}}}{n_{\text{inlet}}}$, where $N$ is the number of particles and $n$ is the number density at the inlet and throat. Previous studies have demonstrated that gas-surface scattering models significantly affect both parameters~\cite{zheng2021design,moon2023performance,ko2023parametric,woodley2024requirements}. This section investigates the effect of surface corrugation on intake performance. Consistent with Section~\ref{sec:5_1}, xenon is used as the working gas, with freestream conditions and VHS parameters taken from Tables~\ref{tab:freestream} and \ref{tab:xenon_vhs}~\cite{emmert2021nrlmsis}. Three scattering models from Table~\ref{tab:scattering_models} are considered: the conventional CLL model and corrugated CLL models with corrugation factors $A = 0.1$ and $0.2$. The intake adopts a conical configuration with a grid duct designed to suppress backscattering and improve $\eta$, as illustrated in Fig.~\ref{fig:intake_geometry}~\cite{rapisarda2023design,zheng2021design,romano2016intake,wu2022recent,moon2023performance}. DSMC simulations are performed in a 3D domain with 145 million particles and 5 million grid cells, using adaptive grid refinement near the surface and a time step of $1 \times 10^{-7}$ s.

\begin{figure}[htb]
  \centering

    \begin{subfigure}[b]{0.48\textwidth}
    \includegraphics[width=\linewidth]{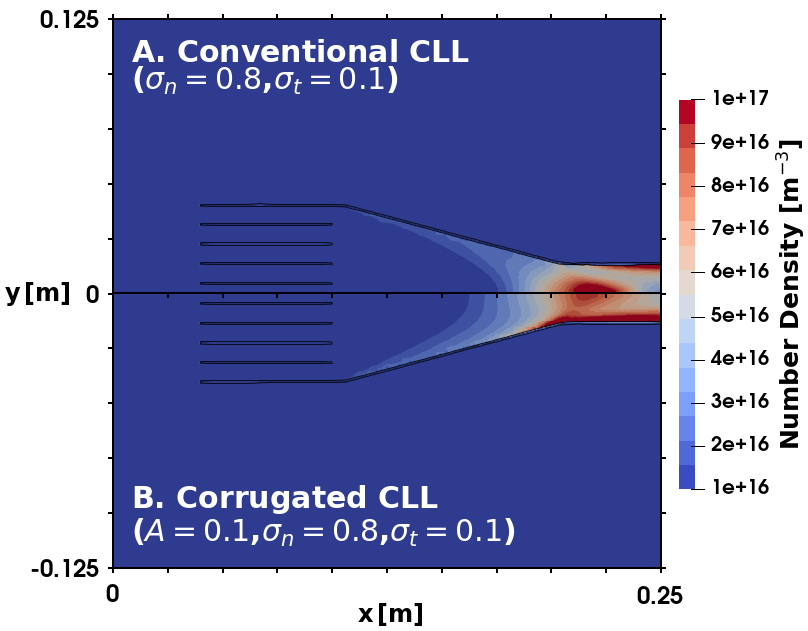}
    \caption{Number density: Cases A-B}
    \label{fig:intake_n_ab}
  \end{subfigure}
  \hfill
  \begin{subfigure}[b]{0.48\textwidth}
    \includegraphics[width=\linewidth]{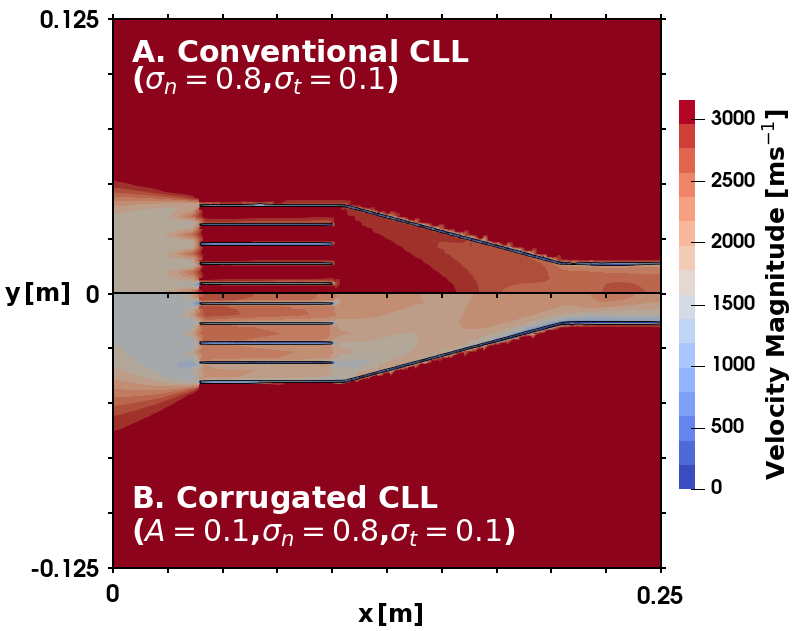}
    \caption{Velocity magnitude: Cases A-B}
    \label{fig:intake_v_ab}
  \end{subfigure}
  \bigskip

    \begin{subfigure}[b]{0.48\textwidth}
    \includegraphics[width=\linewidth]{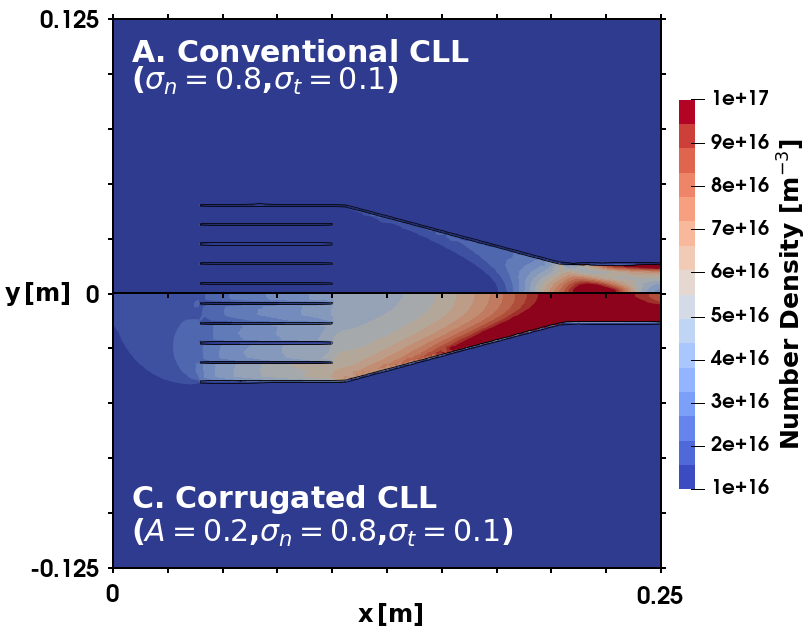}
    \caption{Number density: Cases A-C}
    \label{fig:intake_n_ac}
  \end{subfigure}
  \hfill
  \begin{subfigure}[b]{0.48\textwidth}
    \includegraphics[width=\linewidth]{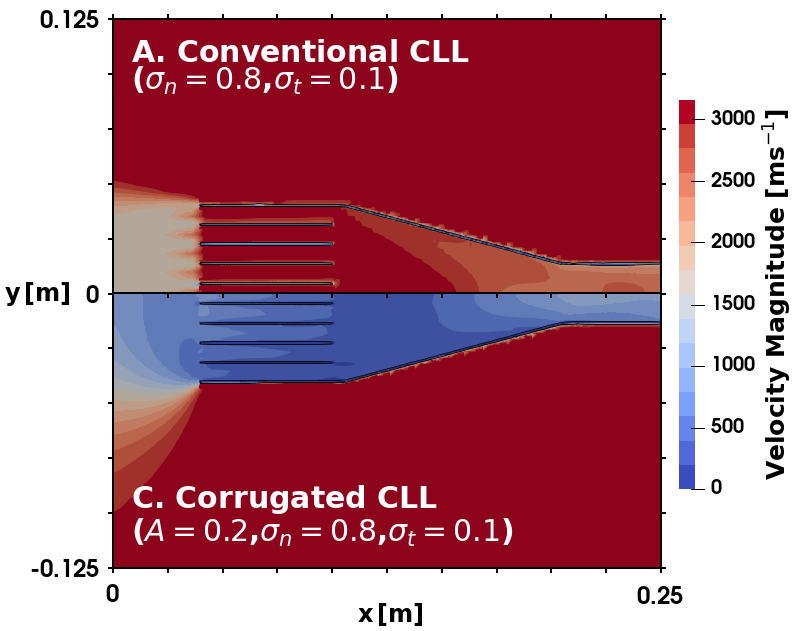}
    \caption{Velocity magnitude: Cases A-C}
    \label{fig:intake_v_ac}
  \end{subfigure}

  \caption{Flow inside an intake: number density and velocity magnitude}
  \label{fig:intake}
\end{figure}

\begin{figure}[ht]
  \centering
  \begin{subfigure}[b]{0.45\textwidth}
    \includegraphics[width=\linewidth]{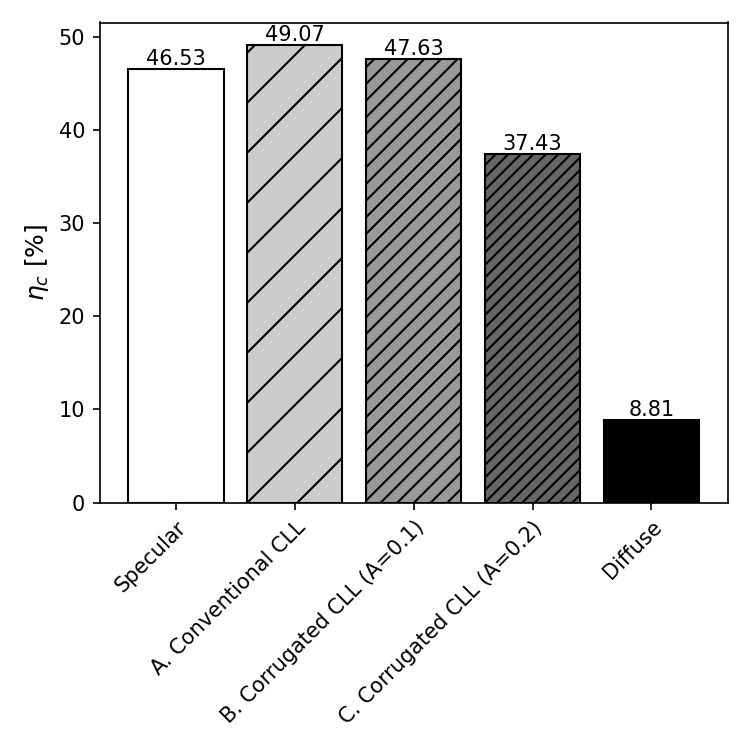}
    \caption{Capture efficiency}
    \label{fig:capture_eff}
  \end{subfigure}
  \quad
  \begin{subfigure}[b]{0.45\textwidth}
    \includegraphics[width=\linewidth]{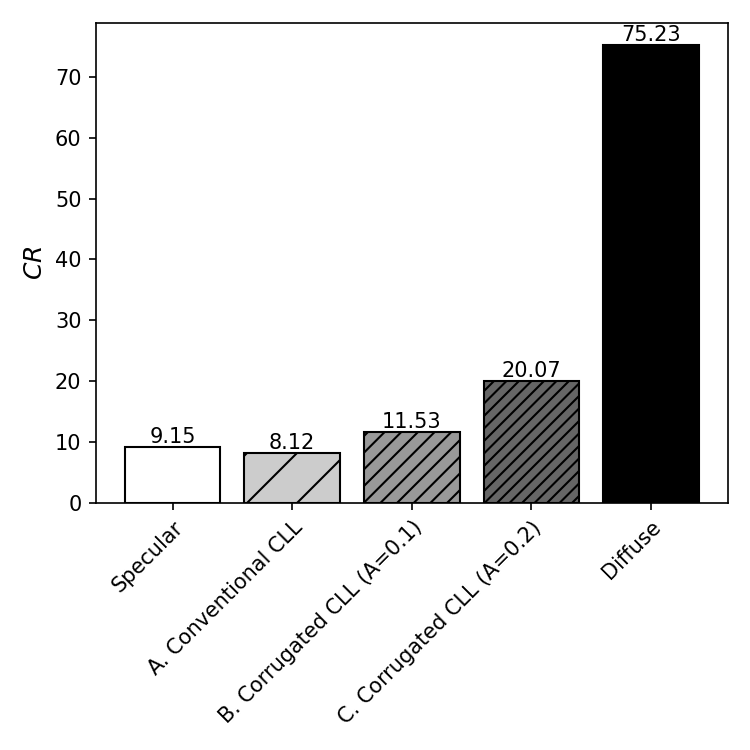}
    \caption{Compression ratio}
    \label{fig:compress_ratio}
  \end{subfigure}

    \caption{Capture efficiency and compression ratio of ducted inlet using different scattering models}
    \label{fig:intake_performance}
\end{figure}


Figure~\ref{fig:intake} shows the number density and velocity magnitude contours within the intake. In all cases, the flow undergoes compression due to the decreasing cross-sectional area, resulting in increased density and reduced velocity. In Case A, density and velocity within the grid duct remain close to freestream values. This is attributed to collisions with the grid duct surface at large incident angles, where the low tangential accommodation coefficient $\sigma_t=0.1$ leads to minimal energy exchange. As the flow encounters the tapered chamber surface, particles reflect at superspecular angles, producing a specular-like focusing effect that generates a sharp density peak along the centerline near the throat. With increasing surface corrugation in Cases B and C, the velocity within the grid duct exhibits greater deceleration, while density increases particularly in Case C. As shown in Fig.~\ref{fig:eac}, the corrugated CLL model exhibits higher total energy accommodation $\sigma$ than the conventional model at an incident angle of 90$^\circ$, which contributes to this behavior. Surface corrugation also increases the thermal component of the reflected velocity, which increases the frequency of gas-surface scattering within the intake. This leads to further momentum and energy accommodation within the intake. The bottom of Fig.~\ref{fig:intake_v_ab}, corresponding to Case B, shows that high-velocity freestream particles undergo structure scattering at the grid duct, while decelerated particles near the throat experience thermal scattering. This indicates a transition from structure to thermal scattering regimes can occur within the intake, highlighting the importance of incorporating surface corrugation into scattering models for accurate characterization of internal flows.

Figure~\ref{fig:intake_performance} presents the intake performance parameters, capture efficiency $\eta$ and compression ratio $CR$. Results for specular and diffuse reflections are also shown, representing the two limiting behaviors of the Maxwell model commonly used in previous intake analyses~\cite{jiang2023aerodynamic,moon2023performance}. Specular reflection maintains the axial momentum of incoming particles, resulting in higher capture efficiency. In contrast, diffuse reflection removes axial momentum, which lowers capture efficiency but increases the compression ratio required for ignition. In Case A, the conventional CLL model's superspecularity increases the axial velocity component. This enhanced axial velocity promotes the particle transport from the inlet to the throat, resulting in higher capture efficiency and a lower compression ratio compared to the specular case. In Cases B and C, surface corrugation increases density and reduces velocity within the intake. These changes lead to decreased capture efficiency and increased compression ratio. Specifically, relative to Case A, Case B yields a 3\% reduction in capture efficiency and a 42\% increase in compression ratio, whereas Case C results in a 23\% reduction in capture efficiency and a 147\% increase in compression ratio. The impact of surface corrugation on intake performance is more pronounced than its effect on drag, as discussed in Section~\ref{sec:5_1}. This is because multiple gas-surface scatterings within the intake amplify the effect of surface corrugation. In VLEO intake flows, where multiple gas–surface interactions occur, accurate performance prediction requires scattering models that incorporate the effects of surface corrugation.


\clearpage
\section{Conclusion} \label{sec:6}
This study investigates the influence of surface corrugation on gas–surface scattering using a corrugated CLL model, which extends the washboard–CLL hybrid approach by incorporating tangential accommodation during local collisions~\cite{liang2018physical}. The corrugated CLL model is validated by reproducing molecular beam scattering experiments~\cite{rettner1991angular}. The corrugated CLL model offers a more accurate representation of structure scattering than the conventional model, predicting broader angular distributions and positively sloped energy distributions. However, it fails to reproduce the experimentally observed increase in angular width with incident energy, as it assumes a fixed level of surface corrugation. When the corrugated CLL model is applied to a macroscopic flow over a cylinder, surface corrugation is found to have limited impact on total drag but alters the ratio between pressure and friction contributions. In intake flows, multiple wall collisions amplify the effect of surface corrugation, increasing compression ratio while reducing capture efficiency. Nevertheless, due to the lack of corrugated CLL model calibration under realistic VLEO conditions, its applicability for modeling gas-surface scattering in VLEO remains limited.
Future work will incorporate energy-dependent surface corrugation, calibrated using high-fidelity simulations or experimental data under realistic VLEO conditions. In addition, the finite-rate surface chemistry framework in DSMC will be extended to account for surface corrugation induced by gas-surface reactions~\cite{gopalan2025development,ko2024mechanism,huh2025drag}. These developments may enable more accurate modeling of gas–surface interactions in rarefied VLEO environments. Furthermore, the proposed corrugated CLL model can be extended to kinetic solvers based on Bhatnagar-Gross-Krook and Fokker–Planck formulations, offering a physically grounded alternative to conventional Maxwell or CLL models~\cite{jun2018assessment,park2024evaluation,kim2024stochastic,kim2024second}.

\section{Acknowledgements}
This material is based upon work supported by the Air Force Office of Scientific Research under award number FA2386-24-1-4074. This work was supported by the National Supercomputing Center with supercomputing resources including technical support(KSC-2025-CRE-0427).

\clearpage

\clearpage

\end{document}